\documentclass[aps,prd,showpacs,superscriptaddress,twocolumn ]{revtex4}

\usepackage{amsfonts}
\usepackage{amsmath}
\usepackage[USenglish]{babel}
\usepackage{graphicx}
\usepackage{color}
\usepackage{latexsym}
\usepackage{subfig}
\usepackage[active]{srcltx}
\usepackage{float}

\DeclareMathAlphabet\mathbfcal{OMS}{cmsy}{b}{n}

\begin{document}

\title{Interaction of a hydrogenlike ion with a planar topological insulator}

\author{A. Mart\'{\i}n-Ruiz}
\email{alberto.martin@nucleares.unam.mx}
\affiliation{Instituto de Ciencias Nucleares, Universidad Nacional Aut\'{o}noma de M\'{e}xico, 04510 M\'{e}xico, Distrito Federal, M\'{e}xico}
\affiliation{{Instituto de Ciencia de Materiales de Madrid, CSIC, Cantoblanco, 28049 Madrid, Spain.}}

\author{L. F. Urrutia}
\email{urrutia@nucleares.unam.mx}
\affiliation{Instituto de Ciencias Nucleares, Universidad Nacional Aut\'{o}noma de M\'{e}xico, 04510 M\'{e}xico, Distrito Federal, M\'{e}xico}

\begin{abstract}
An electric charge near the surface of a topological insulator (TI) induces an image magnetic monopole. Here we study the spectra of hydrogenlike ions near the surface of a planar TI, taking into account the modifications which arise due to the presence of the image monopole magnetic fields. In fact, the atom-TI interaction provides additional contributions to the Casimir-Polder potential while the ion-TI interaction modifies the energy shifts in the spectrum, which now became distance dependent. We show that the hyperfine structure is sensitive to the image magnetic monopole fields in states with nonzero angular momentum, and that circular Rydberg ions can enhance the maximal energy shifts. We discuss in detail the energy splitting of the $n$P$_{1/2}$ and $n$P$_{3/2}$ states in hydrogen. We also analyze the Casimir-Polder potential and find that this magnetic interaction produces a large distance repulsive tail for some particular atomic states. A sizable value of the maximum of the potential requires TIs with very low values of the permittivity together with high values of the topological magnetoelectric polarization.
\end{abstract}

\maketitle

\section{Introduction}

Most quantum states of matter are categorized by the symmetries they break, and they are described by effective Landau-Ginzburg theories \cite{Anderson}. However, topological phases evade traditional symmetry-breaking classification schemes. Instead, in the low-energy limit they are described by topological field theories with quantized coefficients \cite{Qi-PRB}. For instance, the quantum Hall effect is described by the topological Chern-Simons theory in (2+1)D \cite{Zhang}, with coefficient corresponding to the quantized Hall conductance. Recently, topological insulators (TIs) in (3+1)D have attracted great attention in condensed matter physics. These materials display nontrivial topological order and are characterized by a fully insulating bulk together with gapless surface states, which are protected by time-reversal (TR) symmetry \cite{Qi-ReviewTI, Hasan-ReviewTI}. This type of topological behavior was first predicted in graphene \cite{Kane-Mele}. It was subsequently predicted and then observed in alloys and stoichiometric crystals that display strong enough spin-orbit coupling to induce band inversion, such as Bi${}_{1-x}$Sb${}_{x}$ \cite{Exp-TI1, Exp-TI2}, Bi${}_{2}$Se${}_{3}$, Bi${}_{2}$Te${}_{3}$, Sb${}_{2}$Te${}_{3}$ \cite{Exp-TI3, Exp-TI4} and TlBiSe${}_{2}$ \cite{Sato}. These discoveries stimulate further exploration of the exotic properties of the TIs.

The peculiar properties of TIs stem from a nontrivial topology of their band structure, but also novel interesting properties at the macroscopic level emerge when they interact with, for example, electromagnetic fields \cite{Qi-PRB}. The full theory accounting for the electromagnetic response of TIs is given by the standard Maxwell Lagrangian plus the additional term $\mathcal{L} _{\theta} = (\alpha / 4 \pi ^{2}) \theta \mathbf{E} \cdot \mathbf{B}$, where  $\mathbf{E}$ and $\mathbf{B}$ are the electromagnetic fields, $\alpha$ is the fine structure constant, and $\theta$ is an angular variable known in particle physics as the axion angle \cite{Wilczek}. In general, the axion angle $\theta$ is a dynamical field; however, as far as the electromagnetic response of TIs is concerned, it is quantized in odd integer values of $\pi$, i.e. $\theta = \pm (2n+1) \pi$, where $n \in \mathbb{Z} ^{+}$, and it can be viewed as a phenomenological parameter in the sense of an effective Landau-Ginzburg theory.

One of the most striking features of this topological response theory is the topological magnetoelectric effect (TME), which consists of a mixing between the electric $\mathbf{E}$ and magnetic induction $\mathbf{B}$ fields at the surface of the material. This is why, in the condensed matter literature, the axion angle is termed the topological magnetoelectric polarization (TMEP). Among the remarkable consequences of the TME, which  we are concerned here, is the appearance of image magnetic monopoles when a pointlike electric charge is brought near to the surface of a TI. This effect, known as the image magnetic monopole effect, was first derived in Ref. \cite{Qi-Science} using the usual method of images of electromagnetism; however, it has also been obtained using different methods, e.g. by the action of the $SL(2,\mathbb{Z})$ duality group on TIs \cite{Karch} and by Green's function techniques \cite{MCU4}. The existence of these image magnetic monopoles is compatible with the Maxwell equation $\nabla \cdot \mathbf{B}=0$, since the resulting magnetic fields are in fact induced by circulating vortex Hall currents on the surface of the TI, which are sourced by the electric charge next to the interface, rather than by a real pointlike magnetic charge. Other TMEs involving the appearance of image current and charge densities of magnetic monopoles have been predicted \cite{MCU4}. Additional effects due to the TME have been envisioned. For example, when polarized light propagates through a TI surface, of which the surface states has been gapped by TR symmetry breaking, a topological Faraday rotation of $1 \sim 10$ mrad appears, which falls in a small window but within the current experimental reach \cite{Maciejko, Tse1, Tse2, Tse3, Crosse1, Crosse2}. On the other hand, the effects of the topological nontriviality on the Casimir effect has also been considered \cite{MCU3, Grushin-PRL, Grushin-PRB}.

The experimental determination of the TME arising from TIs in (3+1)D has proved to be rather difficult. This is so because there is an important difference between the $\theta $ term for (3+1)D topological insulators and the (2+1)D Chern-Simons term for quantum Hall systems \cite{Qi-ReviewTI}. In (2+1)D, a simple dimensional analysis reveals that the topological Chern-Simons term dominates over the nontopological Maxwell term at low energies. However, in (3+1)D, both terms are equally important at low energies since they have the same scaling dimension. This implies that, for (3+1)D TIs, the topological response always coexist with the ordinary electromagnetic response, thus making the detection of the TME of TIs experimentally challenging. Despite this limitation, it was recently reported the measurement of a universal topological Faraday rotation angle equal to the fine structure constant when linearly polarized radiation passes through two surfaces of the TI HgTe \cite{Dziom}.

In order to motivate our approach to study the TME let us recall that the presence of an atom in front of a material body will modify its quantum properties, such as the magnitude of the energy levels and the decay rates of the excited states, which now  become functions of the distance between the atom and the body. For a given quantum state, the energy of each atomic level can be interpreted as the interaction energy of the system, yielding the Casimir-Polder (CP) potential experienced by the atom in this state. In this way, distance dependent energy levels of an atom can be analyzed from two alternative perspectives, which have been very successful and well studied along the years: (i) the investigation of dispersion forces \cite{BUHMANN,MILTON} and (ii) the consideration of atomic spectroscopy \cite{Spectroscopy}. According to 
Ref. \cite{BUHMANN} we will denote by CP interactions those between an atom and a body. In a first approximation, the CP interaction can be understood as arising from the dipole induced by the polarization of the atom, which interacts with an image dipole inside the material required  to satisfy the boundary conditions at the surface of the body. If the material body is, for example, a topological insulator, additional interactions arise due to the TME: the charges in the atom will also induce image magnetic monopoles inside the TI, which will in turn interact with the electron via the standard minimal coupling. Since the calculation of the nonretarded force on a charge in front of a metallic plate \cite{LennardJones}, followed by its generalization including retardation \cite{CasimirPolder}, the CP interaction has been profusely studied in diverse materials and geometries. Such extensive interest is motivated by the relevance of CP forces in many branches of science like field theory, cosmology, molecular physics, colloid science, biology, astrophysics, micro and nanotechnology, for example. The measurements  of CP forces has also experienced a high degree of sophistication ranging from experiments based upon classical scattering \cite{Raskin,Haroche1}, quantum scattering \cite{Shimizu, Friedrich}, and spectroscopic measurements \cite{Haroche2,Sukenik,Ivanov}. For a detailed account of the theoretical and experimental work on the CP interaction, including the appropriate references, see for example \cite{BUHMANN}.

Within the realm of atomic spectroscopy and because of the well-developed theory together with  a large tradition in high precision measurements, 
hydrogenlike ions
could provide an attractive test bed for studying the TME, since their hyperfine structure
turns out to be sensitive to the image monopole magnetic fields. The case of circular Rydberg ions will be of relevance because they provide an enhancement  of the TME contribution with respect to the optical one.

The specific problem we shall consider is that of an hydrogenlike ion, including the case of a hydrogen atom, near the surface of a TI. The TI is assumed to be covered with a thin magnetic layer to gap the surface states. Due to the TME, the atomic charges  produce image magnetic monopoles inside the TI, whose magnetic fields cause additional small shifts in the energy levels of the ion. For a given state, the corresponding energy  provides the nonretarded CP potential as well as the distance dependent energy shifts. Also, we discuss the spectra of the lines where the new contributions from the TME induced by the topological insulator arise. Since the splitting of the energy levels depends mainly on the ion-surface distance, we focus on the case where: i) there is a negligible wave-function overlap between the electron and the surface states, ii) the ion-TI interaction is dominated by nonretarded electromagnetic forces, and iii) perturbation theory is valid.

The paper is organized as follows. In Sec.\ref{ResponseTIs} we review the basics of the electromagnetic response of TR invariant topological insulators in (3+1)D. The Hamiltonian describing the interaction of the ion with the TI is derived in Sec. \ref{Interaction}. We analyze the order of magnitude of each contribution and we retain the more important ones. The energy shifts of  circular Rydberg hydrogenlike ions are discussed in Sec. \ref{Energy-Levels}, where we consider  the separate  cases where  the ion and the TI are embedded either in the same dielectric  media or in a different  one. The former situation also contributes to the amplification of the TME. Section \ref{Energy-Levels-Hydrogen} includes the calculation of the energy level shifts of the hyperfine spectrum of the hydrogen in the $n$P$_{3/2}$ and $n$P$_{1/2}$ states, which constitute the basis for the analysis in the next section \ref{Casimir-Polder}, where we discuss the resulting Casimir-Polder interaction in each of the previously determined states. A concluding summary of our results and a discussion on the limitations of our model comprises Section \ref{Discussion}. Throughout the paper, Lorentz-Heaviside units are assumed ($\hbar =c=1$), the metric signature will be taken as $(+,-,-,-)$ and $\epsilon ^{0123} = +1$.

\section{Electromagnetic response of (3+1)D topological insulators}

\label{ResponseTIs}

The low-energy effective field theory governing the electromagnetic response of (3+1)D topological insulators, independently of the microscopic details, is defined by the action
\begin{equation}
S = \int \left[ \frac{1}{8\pi} \left( \varepsilon \mathbf{E} ^{2} - \frac{1}{\mu} \mathbf{B} ^{2} \right) + \frac{\alpha}{4 \pi ^{2}} \theta \mathbf{E} \cdot \mathbf{B} \right] d ^{4} x ,  \label{Lagrangian}
\end{equation}
where $\mathbf{E}$ and $\mathbf{B}$ are the electromagnetic fields, $\alpha \simeq 1/137$ is the fine structure constant, $\varepsilon$ and $\mu$ are the permittivity and permeability, respectively, and $\theta$ is the TMEP (axion field). When the theory is defined on a manifold without boundary, TR symmetry indicates that there are only two nonequivalent allowed values of $\theta$ which are $0$ and $\pi$ modulo $2 \pi$. This leads to the $\mathbb{Z} _{2}$ classification of three-dimensional TR invariant TIs. For a manifold with a boundary, TR symmetry is broken even if $\theta = \pi$ (modulo $2 \pi$) in the action (\ref{Lagrangian}), and nontrivial metallic surface states appear. The theory is a fair description of the whole system (bulk $+$ boundary) only when a TR breaking perturbation is induced on the surface to gap the surface states, for instance, by means of a magnetic perturbation (applied field and/or film coating) \cite{Qi-ReviewTI, Hasan-ReviewTI} or even by using commensurate out- and in-plane antiferromagnetic or ferrimagnetic insulating thin films \cite{Laszlo}. These surface states have an anomaly which cancel the TR breaking term, thus restoring the TR symmetry of the whole system. In this situation, $\theta$ is quantized in odd integer values of $\pi $ such that $\theta = \pm ( 2n + 1 ) \pi $, where $2n + 1$ corresponds to the number of Dirac fermions on the surface. In this work we consider that the TR perturbation is a magnetic coating of small thickness, such that the two signs correspond to the two possible orientations of the magnetization in the direction perpendicular to the surface. Physically, the axionic term in Eq. (\ref{Lagrangian}) is generated by a quantized Hall effect on the surface of the TI leading to a quantized magnetoelectric response in units of the fine structure constant.

The electromagnetic response of TIs is still described by the ordinary Maxwell equations 
\begin{align}
\nabla \cdot \mathbf{D}=4\pi \rho \phantom{0}\quad & ,\quad \nabla \times 
\mathbf{H}=\frac{\partial \mathbf{D}}{\partial t}+4\pi \mathbf{J},  \notag \\
\nabla \cdot \mathbf{B}=0\phantom{4 \pi \rho}\quad & ,\quad \nabla \times 
\mathbf{E}=-\frac{\partial \mathbf{B}}{\partial t},  \label{MaxEqs}
\end{align}%
with the modified constitutive relations
\begin{equation}
\mathbf{D}=\varepsilon \mathbf{E}+\frac{\alpha }{\pi }\theta \mathbf{B}%
\qquad ,\qquad \mathbf{H}=\frac{1}{\mu }\mathbf{B}-\frac{\alpha }{\pi }%
\theta \mathbf{E}.  \label{ConsitutiveRel}
\end{equation}
The first term in each constitutive relation is the usual electromagnetic term defined in terms of the permittivity and permeability functions, giving rise to the ordinary electromagnetic phenomena. Interestingly, the second term in each constitutive relation, which arises from the axionic term in Eq. (\ref{Lagrangian}), leads to a mixing between the electric $\mathbf{E}$ and magnetic induction $\mathbf{B}$ fields. Importantly, the quantization of the TMEP depends only on the TR symmetry and the bulk topology; it is therefore universal and independent of any material details, thus  guaranteeing the robustness of the TME.

The general solution to the modified Maxwell equations (\ref{MaxEqs}) in the presence of planar, spherical and cylindrical TIs has been recently elaborated by means of Green's function techniques \cite{MCU1, MCU2, MCU4}. Knowledge of the Green's function allows one to compute the electromagnetic potential $A ^{\mu} = (\phi , \mathbf{A})$ at any point from an arbitrary distribution of sources $J ^{\mu} = (\rho ,\mathbf{J})$ via
\begin{equation}
A ^{\mu} (\mathbf{r}) = \int G _{\phantom{\mu} \nu}^{\mu} (\mathbf{r},\mathbf{r} ^{\prime }) J ^{\nu}(\mathbf{r} ^{\prime} ) d ^{3} \mathbf{r} ^{\prime } , \label{Sol4Potential}
\end{equation}
where the Green's function $G _{\phantom{\mu} \nu}^{\mu}$ contains all the information concerning the geometry and boundary conditions on the surface of the TI. Due to the gauge invariance of the action (\ref{Lagrangian}), the electrostatic and magnetostatic fields are defined in terms of the potential $A ^{\mu}$ according to $\mathbf{E} = -\nabla \phi$ and $\mathbf{B} = \nabla \times \mathbf{A}$, as usual. As can be seen from Eq. (\ref{Sol4Potential}), the nondiagonal components of the Green's function are the responsible of the TME.

\section{Hydrogenlike ion near the surface of a TI}

\label{Interaction}

\begin{figure}[tbp]
\begin{center}
\includegraphics[scale=0.5]{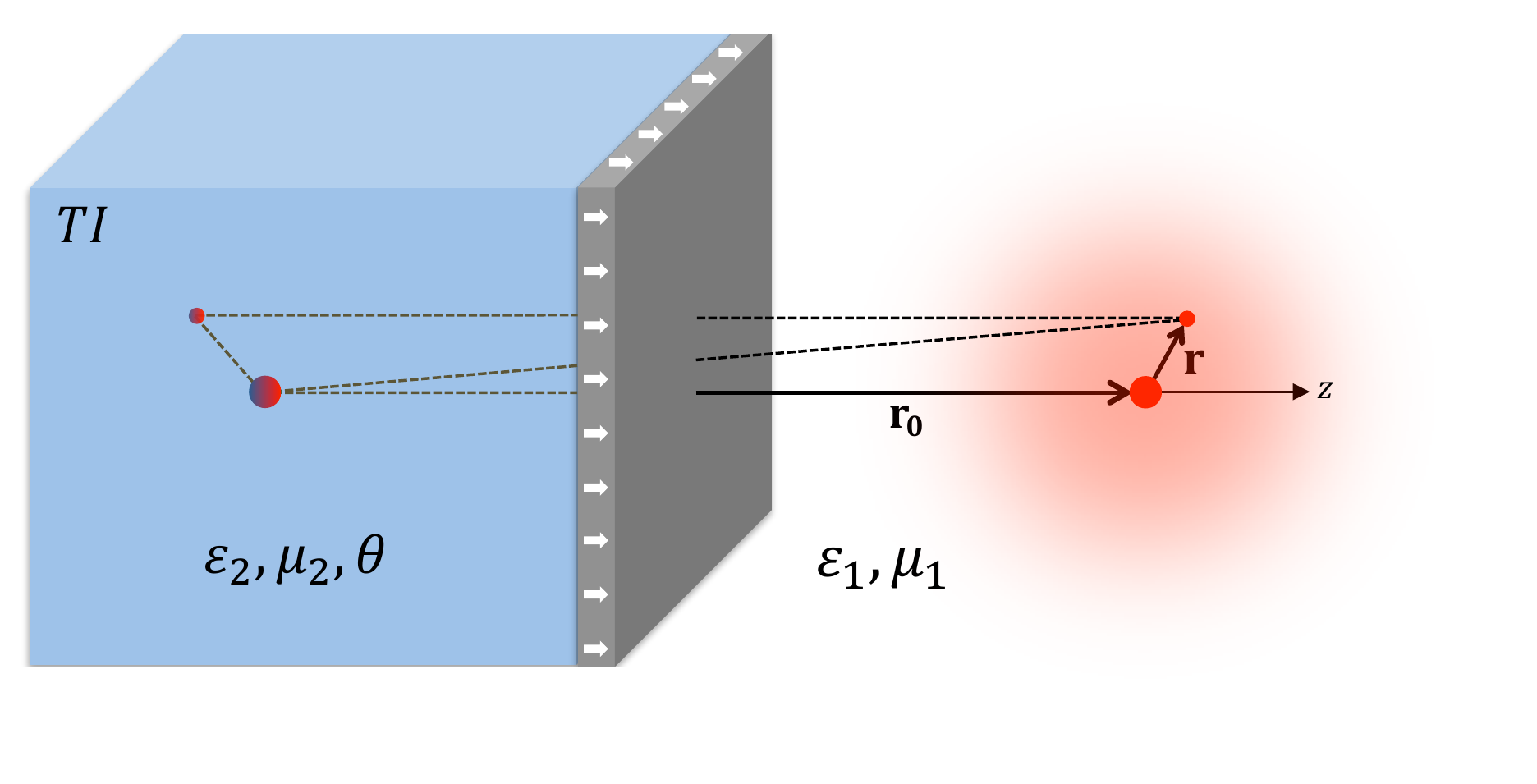}
\end{center}
\caption{{\protect\small An hydrogenlike ion at position $\mathbf{r} _{0}$ near a three-dimensional TI. The TI is covered with a thin magnetic layer of thickness $ w \ll |\mathbf{r} _{0}|$ (not to scale) which controls the sign of the TMEP $\protect\theta$.}}
\label{figure}
\end{figure}

Let us consider an hydrogenlike ion near a three-dimensional TI half-space, as shown in Fig. \ref{figure}, and let us restrict ourselves to the nonretarded approximation. As is well known, this regime is valid for distances $b$ such that $b \ll \lambda_{\rm{C}}$, where $\lambda _{\rm{C}}$ can be estimated as the maximum wavelength characterizing the transitions between the specific energy levels being probed in the ion \cite{BUHMANN,Eberlein}. In what follows we will consider the nucleus to be fixed at $\mathbf{r}_{0} = b \hat{\textbf{e}}_{z}$, and we assume that the TI is covered with a thin magnetic layer of thickness $w \ll b$ and magnetization $\mathbf{M} = M \hat{\textbf{e}}_{z}$, such that there is a negligible wave-function overlap between the atomic electron and the surface states. Thus, we henceforth assume that $w \ll b \ll \lambda _{\rm{C}}$. We stress here that the only effect of the magnetic coating is to gap the surface states. However, the ferromagnet makes a magnetic field and the energy shifts of the atomic spectrum are to be measured as a function of the magnetization $M$. The effects we shall discuss in the following are defined as the linear extrapolation of the energy shifts as $M \rightarrow 0^{+}$, in which limit the nontopological contributions are removed. In Section \ref{Discussion} we will discuss the effects of the magnetic coating on the energy levels in more detail.

In the nonretarded regime, the CP interaction between two atoms (two hydrogen atoms, for instance) is achieved by computing the Coulomb interaction between all charges of one atom and all charges of the other. Since in our model we have to take into account many pairwise Coulomb interactions (between the charges of the atom and their electric images), it is convenient to introduce the Coulomb interaction Hamiltonian, 
\begin{equation}
U _{\mbox{\scriptsize Coul}} = \frac{1}{2} \int \rho (\mathbf{r}) G _{\phantom{0}0} ^{0} (\mathbf{r},\mathbf{r} ^{\prime }) \rho (\mathbf{r}^{\prime }) d ^{3} \mathbf{r} d ^{3} \mathbf{r} ^{\prime} ,  \label{U-Coulomb}
\end{equation}
where $\rho$ includes the two opposite charges of the ion.

Due to the TME, the atomic charges will also produce image magnetic monopoles located inside the TI, whose magnetic fields will in turn interact with the atomic electron. This interaction results from the coupling of the magnetic field produced by the image monopoles with the electron spin and their orbital motion. Therefore, the Hamiltonian we consider is 
\begin{equation}
H = \frac{{\boldsymbol{\Pi}}^2}{2m _{e}}  + U _{\mbox{\scriptsize Coul}} +\frac{e}{m _{e}} \mathbf{S} \cdot \mathbf{B} + V _{\mbox{\scriptsize fs}} + V _{\mbox{\scriptsize hfs}},  \label{Hamiltonian}
\end{equation}
where $m _{e}$ is the electron mass, $e > 0 $ is the magnitude of the electron charge, $\boldsymbol{\Pi} = \mathbf{p} + e \mathbf{A}$ is the canonical momentum and $\mathbf{B} = \nabla \times \mathbf{A}$ is the magnetic field associated with the image magnetic monopoles, one contribution arising from the nucleus and the other from the electron. Besides, $U _{\mbox{\scriptsize Coul}}$ is the Coulomb interaction energy (\ref{U-Coulomb}), $V _{\mbox{\scriptsize fs}}$ are the fine structure contributions (relativistic energy correction, spin-orbit coupling and Darwin term) and $V_{\mbox{\scriptsize hfs}}$ is the hyperfine interaction.  

In our Hamiltonian, we have considered those terms which give the most important contributions to the energy spectrum. However, there are other smaller terms, such as the interaction between the image monopole magnetic fields and the nuclear spin, and also the interaction between the image electric dipole produced by the atomic magnetic moment and the atomic electric dipole. Next let us study each term separately.

Since the ion-surface distance, though small to make retardation effects negligible, is much greater than the Bohr radius $a _{0}$, a Taylor expansion in the electromagnetic fields can be performed \cite{Eberlein, Eberlein2, Souza}. Then, treating the ion as an electric composite system, the effective charge density can be written as
\begin{equation}
\rho (\mathbf{r} ^{\prime }) = e \left[ Z \delta (\mathbf{r}^{\prime } - \mathbf{r} _{0}) - \delta (\mathbf{r} ^{\prime} - \mathbf{r} _{0} -\mathbf{r}) \right] , \label{RHO}
\end{equation}
where $Z$ is the atomic number, and $\textbf{r}$ is the vector which localizes the electron from the nucleus.
It is convenient to split the $00$-component of the Green's function as \cite{MCU4}
\begin{equation}
G _{\phantom{0}0} ^{0} (\mathbf{r} , \mathbf{r} ^{\prime }) = G( \mathbf{r} , \mathbf{r} ^{\prime}) + G _{S} (\mathbf{r} , \mathbf{r} ^{\prime}) , \label{G00TOTAL}
\end{equation}
where the first term 
\begin{equation}
G (\mathbf{r} , \mathbf{r} ^{\prime }) =  \frac{1}{\sqrt{(x - x ^{\prime}) ^{2} + ( y - y ^{\prime}) ^{2} + ( z - z ^ {\prime}) ^{2}}}
\label{G00FREE}
\end{equation} 
is the Green's function in unbounded space. The second term 
\begin{equation}
G _{S} (\mathbf{r},\mathbf{r}^{\prime })  = \frac{\kappa}{\sqrt{( x - x ^{\prime}) ^{2} + ( y - y ^{\prime}) ^{2} + ( z + z ^{\prime}) ^{2}}},
\label{G00H}
\end{equation}
with
\begin{equation}
\kappa = \frac{1}{\varepsilon _{1}} \frac{(\varepsilon _{1} - \varepsilon _{2})( 1 / \mu _{1} + 1 / \mu _{2} ) - \tilde{\theta}^{2}}{(\varepsilon _{1} + \varepsilon _{2})(1 / \mu _{1} + 1 / \mu _{2}) + \tilde{\theta}^{2}} \quad ,\quad \tilde{\theta}\equiv \alpha \frac{\theta}{\pi},  \label{kappa}
\end{equation}
is a solution of the homogeneous Laplace equation, such that $G_{\phantom{0}0}^{0} (\mathbf{r},\mathbf{r}^{\prime })$ satisfies the required boundary conditions at the surface of the TI \cite{MCU3}. Let us emphasize that $\tilde{\theta}$ in Eq. (\ref{kappa}) is of order $\alpha$. Using the aforementioned charge distribution and the Green's functions, the Coulomb interaction (\ref{U-Coulomb}) becomes
\begin{align}
U _{\mbox{\scriptsize Coul}} =& \frac{e ^{2}}{2} \Big[ Z ^{2} G (\mathbf{r}_{0},\mathbf{r} _{0}) - 2 Z G (\mathbf{r} _{0} + \mathbf{r} , \mathbf{r} _{0}) + G (\mathbf{r} _{0} + \mathbf{r} , \mathbf{r} _{0} +\mathbf{r}) \Big] \notag \\ & + \frac{e ^{2}}{2} \Big[ Z ^{2} G _{S} (\mathbf{r} _{0} , \mathbf{r} _{0}) + G _{S} (\mathbf{r} _{0} + \textbf{r} , \mathbf{r} _{0} + \textbf{r}) \notag \\ & \hspace{0.5cm} - Z G _{S}(\mathbf{r} _{0} , \mathbf{r} _{0} + \mathbf{r}) - Z G _{S} (\mathbf{r} _{0} + \mathbf{r},\mathbf{r} _{0}) \Big] .  \label{U-Coulomb2}
\end{align}
The terms $Z ^{2} G (\mathbf{r} _{0} , \mathbf{r} _{0})$ and $G (\mathbf{r} _{0} + \mathbf{r} , \mathbf{r} _{0} + \mathbf{r})$ in this expression are the divergent self-energies of  the nucleus  and the electron, respectively, which we discard. The term $Z G (\mathbf{r} _{0} , \mathbf{r} _{0} + \mathbf{r})$ corresponds to the nucleus-electron interaction. The contributions due to the presence of the TI are given by the remaining four terms and the Coulomb energy now takes the form 
\begin{align}
U _{\mbox{\scriptsize Coul}} =& - \frac{Z e^{2}}{\varepsilon _{1} r} + \frac{e ^{2}}{2} \Big[ Z ^{2} G _{S} (\mathbf{r} _{0} , \mathbf{r} _{0}) - Z G _{S} (\mathbf{r} _{0} + \mathbf{r},\mathbf{r} _{0}) \notag \\ & - Z G _{S}(\mathbf{r} _{0} , \mathbf{r} _{0} + \mathbf{r}) + G _{S} (\mathbf{r} _{0} + \textbf{r} , \mathbf{r} _{0} + \textbf{r}) \Big] .  \label{U-Coulomb21}
\end{align}
In the limit $\vert \mathbf{r} \vert \ll \vert \mathbf{r}_{0} \vert$, when the dimensions of the ion are small compared with the nucleus-interface distance $\vert \mathbf{r} _{0} \vert$, the additional  terms can be written as derivatives of $G _{S}$. Making a Taylor expansion of $G _{S}(\mathbf{r}_{0} , \mathbf{r} _{0} + \mathbf{r})$ in powers of $\mathbf{r}$ up to second order produces
\begin{align}
G _{S} ^{(2)} (\mathbf{r} _{0} , \mathbf{r}_{0} +\mathbf{r}) \simeq & G _{S} (\mathbf{r} _{0},\mathbf{r} _{0}) + \left( \mathbf{r} \cdot \nabla ^{\prime} \right) G _{S} (\mathbf{r} _{0},\mathbf{r}^{\prime}) \vert _{\mathbf{r} ^{\prime} = \mathbf{r}_{0}} \notag \\ & + \frac{1}{2} \left( \mathbf{r} \cdot \nabla ^{\prime} \right) ^{2} G _{S} (\mathbf{r} _{0},\mathbf{r}^{\prime}) \vert _{\mathbf{r} ^{\prime} = \mathbf{r}_{0}} . \label{Appox1}
\end{align}
Analogously, a Taylor expansion for $G _{S}(\mathbf{r}_{0} + \mathbf{r} , \mathbf{r} _{0})$ yields
\begin{align}
G _{S} ^{(2)} (\mathbf{r} _{0} +\mathbf{r} , \mathbf{r}_{0} ) \simeq & G _{S} (\mathbf{r} _{0},\mathbf{r} _{0}) + \left( \mathbf{r} \cdot \nabla ^{\prime} \right) G _{S} (\mathbf{r} ^{\prime} , \mathbf{r} _{0}) \vert _{\mathbf{r} ^{\prime} = \mathbf{r}_{0}} \notag \\ & + \frac{1}{2} \left( \mathbf{r} \cdot \nabla ^{\prime} \right) ^{2} G _{S} (\mathbf{r} ^{\prime} , \mathbf{r} _{0}) \vert _{\mathbf{r} ^{\prime} = \mathbf{r}_{0}} . \label{Appox2}
\end{align}
Using the previous results one can further establish
\begin{align}
& G _{S} ^{(2)} (\textbf{r} _{0} + \textbf{r} , \textbf{r} _{0} + \textbf{r}) + G _{S} (\textbf{r} _{0} , \textbf{r} _{0}) \phantom{\frac{1}{2}} \notag \\ & \hspace{1.5cm} \simeq G _{S} ^{(2)} (\mathbf{r} _{0} , \mathbf{r}_{0} +\mathbf{r}) + G _{S} ^{(2)} (\mathbf{r} _{0} + \mathbf{r} , \mathbf{r}_{0}) \notag \\ & \hspace{1.5cm} \phantom{\frac{1}{2}} + (\textbf{r} \cdot \nabla ^{\prime}) (\textbf{r} \cdot \nabla ^{\prime \prime}) G _{S} (\textbf{r} ^{\prime} , \textbf{r} ^{\prime \prime}) \vert _{\textbf{r} ^{\prime \prime} = \textbf{r} ^{\prime} = \textbf{r} _{0} } .  \label{Appox3}
\end{align}
Substituting the expressions (\ref{Appox1}), (\ref{Appox2}) and (\ref{Appox3}) in Eq. (\ref{U-Coulomb21}) yields our final expression for the Coulomb energy
\begin{align}
U _{\mbox{\scriptsize Coul}} =& - \frac{Z e ^{2}}{\varepsilon _{1} r} + \frac{(Z - 1) ^{2} e ^{2}}{2} G _{S} (\textbf{r} _{0} , \textbf{r} _{0}) \notag \\ &+ (1 - Z) e ^{2} \Big[ (\textbf{r} \cdot \nabla ^{\prime}) G _{S} (\textbf{r} _{0} , \textbf{r} ^{\prime}) \Big| _{\textbf{r} ^{\prime} = \textbf{r} _{0}} \notag \\ & + \frac{1}{2} (\textbf{r} \cdot \nabla ^{\prime}) ^{2} G _{S} (\textbf{r} _{0} , \textbf{r} ^{\prime}) \Big| _{\textbf{r} ^{\prime} = \textbf{r} _{0}} \Big] \notag \\ & + \frac{e ^{2}}{2} (\textbf{r} \cdot \nabla ^{\prime}) (\textbf{r} \cdot \nabla ^{\prime \prime}) G _{S} (\textbf{r} ^{\prime} , \textbf{r} ^{\prime \prime}) \Big| _{\textbf{r} ^{\prime \prime} = \textbf{r} ^{\prime} = \textbf{r} _{0}}  . \label{U-Coulomb3}
\end{align}
Using the Green's function $G_{S}$ defined above, one obtains 
\begin{gather}
G _{S} (\textbf{r} _{0} , \textbf{r} _{0}) = \frac{\kappa}{2 b} \quad , \quad (\textbf{r} \cdot \nabla ^{\prime \prime}) G _{S} (\textbf{r} _{0} , \textbf{r} ^{\prime \prime}) \vert _{\textbf{r} ^{\prime \prime} = \textbf{r} _{0}} = - \frac{\kappa z}{4 b ^{2}} , \notag \\ (\textbf{r} \cdot \nabla ^{\prime}) (\textbf{r} \cdot \nabla ^{\prime}) G _{S} (\textbf{r} ^{\prime} , \textbf{r} _{0}) \vert _{\textbf{r} ^{\prime} = \textbf{r} _{0} } = \kappa \frac{3 z ^{2} - r ^{2}}{8 b ^{3}} , \notag \\ (\textbf{r} \cdot \nabla ^{\prime}) (\textbf{r} \cdot \nabla ^{\prime \prime}) G _{S} (\textbf{r} ^{\prime} , \textbf{r} ^{\prime \prime}) \vert _{\textbf{r} ^{\prime \prime} = + \textbf{r} ^{\prime} = \textbf{r} _{0} } = \kappa \frac{r ^{2} + z ^{2}}{8 b ^{3}} , \label{Identity}
\end{gather}
such that the Coulomb interaction simplifies to 
\begin{equation}
U _{\mbox{\scriptsize Coul}} = - \frac{Z e^{2}}{\varepsilon _{1} r} + \delta U _{0}  + \delta U _{1} + \delta U _{2} .  \label{Coulomb-Energy}
\end{equation}
The first term is the usual Coulomb interaction experienced by the atomic electron due to the nucleus. The second term, $\delta U _{0} = \kappa (Z - 1) ^{2} (e ^{2}/4b)$, which corresponds to the interaction between the effective atomic charge and its own image, does not depend on the electron coordinates and thus it is not considered for the purposes of this paper. The last terms 
\begin{gather}
\delta U _{1} = \frac{\kappa (Z-1) e ^{2}}{4 b ^{2}} z \quad , \quad \delta U _{2} = \frac{\kappa e ^{2}}{16 b ^{3}} \left[ Z r ^{2} - (3Z - 4) z ^{2} \right], \label{OpticalTerm}
\end{gather}
constitute the optical contribution to the attractive CP interaction due to the presence of the TI.

Now let us consider the new terms which are the direct manifestation of the image monopole magnetic fields, i.e., $(e / m _{e}) \textbf{A} \cdot \textbf{p}$ and $(e ^{2} / 2 m _{e}) \textbf{A} ^{2}$. These terms will provide additional corrections to the standard CP interaction arising from the presence of the TI. In terms of the Green's function $G_{\phantom{i}0}^{i}(\mathbf{x},\mathbf{x} ^{\prime })$ \cite{MCU4} the vector
potential is
\begin{equation}
A ^{i} (\mathbf{x}) = \int G _{\phantom{i} 0} ^{i}(\mathbf{x},\mathbf{x} ^{\prime}) \rho (\mathbf{x}^{\prime }) d ^{3} \mathbf{x} ^{\prime} ,
\end{equation} 
where $\rho$ is the previously defined charge density in Eq. (\ref{RHO}). This yields
\begin{align}
A ^{i} (\textbf{x}) = Z e G ^{i} _{\phantom{i}0} (\textbf{x}, \textbf{r} _{0}) - e G ^{i} _{\phantom{i}0} (\textbf{x}, \textbf{r} _{0} + \textbf{r}). \label{VectorPotential20}
\end{align}
In the coordinate system attached to the nucleus the vector potencial becomes
\begin{align}
A ^{i} (\textbf{r}) = Z e G ^{i} _{\phantom{i}0} (\textbf{r} _{0} + \textbf{r}, \textbf{r} _{0}) - e G ^{i} _{\phantom{i}0} (\textbf{r} _{0} + \textbf{r} , \textbf{r} _{0} + \textbf{r}) . \label{VectorPotential2}
\end{align}
The first term corresponds to the vector potential produced by the image monopole of the nucleus on the electron, while the second term is the vector potential of the image monopole of the electron on the electron itself.
The calculation  of the vector potential (\ref{VectorPotential2}) starts from the 
 components of the Green's function
\begin{align}
e G ^{i} _{\phantom{i}0} (\textbf{x} , \textbf{x} ^{\prime}) = \frac{g \epsilon ^{0ji3} R _{j}}{R ^{2}} \left( 1 - \frac{z + z ^{\prime}}{\sqrt{R ^{2} + (z + z ^{\prime}) ^{2}}} \right) , \label{Green-i0}
\end{align}
where $R ^{j} = (x - x ^{\prime}) \hat{\textbf{e}} _{x} + (y - y ^{\prime}) \hat{\textbf{e}} _{y}$ and
\begin{align}
g = \frac{2 e \tilde{\theta}}{(\varepsilon _{1} + \varepsilon _{2}) (1 / \mu _{1} + 1 / \mu _{2}) + \tilde{\theta} ^{2}}
\label{etheta}
\end{align}
is the magnitude of the image magnetic monopole of an electric charge $e$.
From Eq. (\ref{Green-i0}) we observe that the expression $G ^{i} _{\phantom{i}0} (\textbf{x} , \textbf{x})$  is ill-defined at a first glance. This calls for a careful determination of the limit $\textbf{x} \rightarrow \textbf{x}^\prime$. A Taylor expansion  of the term in round brackets in the right hand side of Eq. (\ref{Green-i0}) leads to 
\begin{align}
e G ^{i} _{\phantom{i}0} (\textbf{x} , \textbf{x} ^{\prime}) \simeq \frac{g \epsilon ^{0ji3} R _{j}}{2} \left[ \frac{1}{(z + z ^{\prime}) ^{2}} + \mathcal{O} \left( \frac{R ^{2(n-1)}}{(z +z ^{\prime}) ^{2n}} \right) \right] . \label{Green-i0-3}
\end{align}
Taking the limit  $\textbf{x} ^{\prime} \rightarrow \textbf{x}$ we obtain that  $R _{j} \rightarrow 0$ and  $z + z ^{\prime} = 2 z$, in such a way that  $G ^{i} _{\phantom{i}0} (\textbf{x} , \textbf{x}) = 0$. Then, Eq. (\ref{VectorPotential2}) reduces to
\begin{align}
A ^{i} (\textbf{r}) = Z e G ^{i} _{\phantom{i}0} (\textbf{r} _{0} + \textbf{r}, \textbf{r} _{0}) . \label{VectorPotential3}
\end{align}
Since  $\vert \textbf{r} \vert \ll \vert \textbf{r} _{0} \vert$, we perform a Taylor expansion up to third order in $\textbf{r}$ to obtain
\begin{align}
A ^{i} (\textbf{r})& \simeq Z e \Big[ G ^{i} _{\phantom{i}0} (\textbf{r} _{0} , \textbf{r} _{0}) + (\textbf{r} \cdot \nabla ^{\prime}) G ^{i} _{\phantom{i}0} (\textbf{x} ^{\prime} , \textbf{r} _{0}) \notag \\& + \frac{1}{2!} (\textbf{r} \cdot \nabla ^{\prime}) ^{2} G ^{i} _{\phantom{i}0} (\textbf{x} ^{\prime} , \textbf{r} _{0}) + \frac{1}{3!} (\textbf{r} \cdot \nabla ^{\prime}) ^{3} G ^{i} _{\phantom{i}0} (\textbf{x} ^{\prime} , \textbf{r} _{0}) \Big] \Big|  _{\textbf{x} ^{\prime} = \textbf{r} _{0}} . \label{VectorPotential4}
\end{align}
The first term in the right hand side of the above equation is zero and the subsequent contributions require also an accurate calculation of the corresponding limit. Using the Green's function $G ^{i} _{\phantom{i}0}$ defined above, one obtains
\begin{align}
(\textbf{r} \cdot \nabla ^{\prime}) G ^{i} _{\phantom{i}0} (\textbf{x} ^{\prime} , \textbf{r} _{0}) \vert _{\textbf{x} ^{\prime} = \textbf{r} _{0}} &= - \frac{g}{8 b ^{2}} \epsilon ^{0ij3} r _{j} , \notag \\ (\textbf{r} \cdot \nabla ^{\prime}) ^{2} G ^{i} _{\phantom{i}0} (\textbf{x} ^{\prime} , \textbf{r} _{0}) \vert _{\textbf{x} ^{\prime} = \textbf{r} _{0}} &= \frac{g z}{4 b ^{3}} \epsilon ^{0ij3} r _{j} , \notag \\ (\textbf{r} \cdot \nabla ^{\prime}) ^{3} G ^{i} _{\phantom{i}0} (\textbf{x} ^{\prime} , \textbf{r} _{0}) \vert _{\textbf{x} ^{\prime} = \textbf{r} _{0}} &= \frac{9 g}{64 b ^{4}} \left( r ^{2} - 5 z ^{2} \right) \epsilon ^{0ij3} r _{j} . \label{Identity2}
\end{align}
In this way, the final expression for the vector potential is
\begin{align}
A ^{i} (\textbf{r}) = - \frac{Z g e}{8 b ^{2}} \left( 1 - \frac{z}{b} - \frac{3}{16} \frac{r ^{2} - 5 z ^{2}}{b ^{2}} \right) \epsilon ^{0ij3} r _{j} . \label{VectorPotential5}
\end{align}
The final contribution of the term $\textbf{A} \cdot {\textbf{p}}$ to the Hamiltonian is
\begin{align}
\frac{e}{m _{e}} \textbf{A} \cdot {\textbf{p}} = \delta V _{\theta 1} + \delta V _{\theta 2} + \delta V _{\theta 3}, \label{A cdot p}
\end{align}
with
\begin{gather}
\delta V _{\theta 1} = - \frac{Z g e}{8 m _{e} b ^{2}} L _{z} \quad , \quad \delta V _{\theta 2} = - \frac{z}{b} \delta V _{\theta 1},  \notag \\ \delta V _{\theta 3} = - \frac{3}{16} \frac{r ^{2} - 5 z ^{2}}{b ^{2}} \delta V _{\theta 1} , \label{A cdot p 2} 
\end{gather}
where $L _{z} = x p _{y} - y p _{x}$ is the  $z$ component of the angular momentum operator.

Next we deal with the quadratic term in the vector potential appearing in the Hamiltonian (\ref{Hamiltonian}). As shown in the expression (\ref{OpticalTerm}) for the contribution $\delta U$, we are considering corrections up the quadratic order in the electron coordinates. Using the corresponding vector potential (\ref{VectorPotential5}) we thus find
 \begin{align}
\frac{e ^{2}}{2 m _{e}} \textbf{A} ^{2} = \frac{\left( Z g e \right) ^{2}}{128 m _{e} b ^{4}} \left( r ^{2} - z ^{2} \right) . \label{A cdot A}
\end{align}
A similar analysis is next performed for the magnetic interaction $(e / m _{e}) \mathbf{S} \cdot \mathbf{B}$, where the magnetic field $\mathbf{B}$ is produced by the image monopoles of the nucleus and the electron. The nucleus is located  at $\textbf{r} _{0} = b \hat{\textbf{e}} _{z}$ and produces an image monopole of magnitude $Zg$ at  $- \textbf{r} _{0}$. In our coordinate system, the magnetic field of such image monopole acting on the electron is 
\begin{align}
\textbf{B} _{n} = Z g \frac{\textbf{r} + 2 \textbf{r} _{0}}{\vert \textbf{r} + 2 \textbf{r} _{0} \vert ^{3}} = Zg \frac{x \hat{\textbf{e}} _{x} + y \hat{\textbf{e}} _{y} + (z + 2b) \hat{\textbf{e}} _{z}}{\left[ R ^{2} + (z+2b) ^{2} \right] ^{3/2}} .
\end{align}
The other contribution to the image monopole magnetic field comes from the electron itself and it is located at ${\mathbf{r}^\prime= x \hat{\textbf{e}}_{x} + y \hat{\textbf{e}}_{y}-(z+b)\hat{\textbf{e}}_{z}}$ with magnitude $-g$. It is given by
\begin{align}
\textbf{B} _{e} = - g \frac{\hat{\textbf{e}} _{z}}{4 (z + b ) ^{2}} . 
\end{align}
The total magnetic field that feels the electron is then
\begin{align}
\textbf{B} = Z g \frac{x \hat{\textbf{e}} _{x} + y \hat{\textbf{e}} _{y} + (z + 2b) \hat{\textbf{e}} _{z}}{\left[ R ^{2} + (z+2b) ^{2} \right] ^{3/2}} - g \frac{\hat{\textbf{e}} _{z}}{4 (z + b ) ^{2}} . \label{MonopolarMagneticFields}
\end{align}
Performing a Taylor expansion up to quadratic order in  $\textbf{r}$, the magnetic interaction thus takes the final form
\begin{align}
\frac{e}{m _{e}} \textbf{S} \cdot \textbf{B} = \delta W _{\theta 1} + \delta W _{\theta 2} ,
\end{align}
where
\begin{align}
\delta W _{\theta 1} &= \frac{ge}{4 m _{e} b ^{2}} (Z-1) S _{z} , \notag \\ \delta W _{\theta 2} &= \frac{g e}{4 m _{e} b ^{3}} \left[ \frac{Z}{2} \left( x S _{x} + y S _{y} \right) - (Z - 2) z S _{z} \right] .  \label{DeltaW}
\end{align}
Having  taken into account each of the previous  contributions to their lowest order in $x _{i} / b ^{2}$, the Hamiltonian (\ref{Hamiltonian}) reduces to
\begin{align}
H = H _{0} &+ \frac{\kappa (Z-1) e ^{2}}{4 b ^{2}} z + \frac{\kappa e ^{2}}{16 b ^{3}} \left[ Z r ^{2} - (3Z - 4) z ^{2} \right] \notag \\ &- \frac{Z g e}{8 m _{e} b ^{2}} \left( 1 - \frac{z}{b} - \frac{3}{16} \frac{r ^{2} - 5 z ^{2}}{b ^{2}} \right) {\textbf{L}} _{z} \notag \\ &+ \frac{\left( Z g e \right) ^{2}}{128 m _{e} b ^{4}} \left( r ^{2} - z ^{2} \right) + \frac{Z g e}{8 m _{e} b ^{3}} \left( x S _{x} + y S _{y} \right) \notag \\ &+ \frac{ge}{4 m _{e} b ^{2}} \left[ (Z-1) - (Z - 2) \frac{z}{b} \right] S _{z} \label{FULLHAM}
\end{align}
where 
\begin{equation}
H _{0} = \frac{\mathbf{p}^{2}}{2m_{e}} - \frac{ Z e ^{2}}{\varepsilon _{1} r} + H _{\rm{fs}} + H _{\rm{hfs}},
\label{H0}
\end{equation}%
is the Hamiltonian of the hydrogenlike ion including fine ($H_{\rm fs}$) and hyperfine ($H_{\rm hfs}$) corrections.

Our next step is to have an estimation of  the relative weights of the different contributions to the mean value of the Hamiltonian (\ref{FULLHAM}). The exact mean values for a given specific atomic levels will be presented in the next sections. The contributions to the CP potential of the optical terms in the right hand side (rhs) of Eq. (\ref{FULLHAM}) can be estimated as
\begin{align}
\frac{\left< \delta U _{1} \right>}{E _{g}} = & - \frac{\kappa \varepsilon _{1}}{2} \frac{Z-1}{Z} \xi ^{2} \left< \cos \vartheta \right> , \\ \frac{\left< \delta U _{2} \right>}{E _{g}} = & - \frac{\kappa}{8} \left( \frac{\varepsilon _{1}}{Z} \right) ^{2} \xi ^{3} \left[ Z - (3Z-4) \left< \cos ^{2} \vartheta \right> \right] ,
\end{align}
where   $\xi = a _{0} /b \ll 1$ and  $E _{g} = - m _{e} \alpha ^{2} / 2 = - 13.6$eV is the ground state energy of the hydrogen atom. For $b$ of the order of $\mu$m we find $\xi \sim \alpha ^{2} \approx 10 ^{-5}$, and therefore we expect $\left< \delta U _{1} \right> \sim 2 \kappa \varepsilon _{1} \times 10 ^{-8}$ eV for $Z \neq 1$, and a null value  for $Z = 1$. In a similar fashion we expect $\left< \delta U _{2} \right> \sim 2 \kappa (\varepsilon _{1} / Z ) ^{2} \times 10 ^{-12}$ eV. We observe that, although these contributions depend crucially on the values of $\kappa$, $\varepsilon _{1}$ and $Z$, they are smaller than the hyperfine structure of the hydrogenlike ion $E _{\mbox{\scriptsize hfs}} \sim (Z/n \varepsilon_1)^3 \times 10^{-7}$ eV and for this reason we take $H _{0}$ in Eq. (\ref{H0}) as our unperturbed system.

The next term in Eq. (\ref{FULLHAM})
arises from the interaction $(e / m _{e}) \textbf{A} \cdot  {\textbf{p}}$. A direct estimation shows that only the first term in the rhs of Eq. (\ref{A cdot p}), which is of the order of 
\begin{align}
\frac{\left< \delta V _{\theta 1} \right>}{E _{g}} = \frac{Z}{2} \alpha ^{2} \xi ^{2} \frac{(\theta / \pi) \left< L _{z} \right> }{(\varepsilon _{1} + \varepsilon _{2})(1 / \mu _{1} + 1 / \mu _{2}) + \tilde{\theta} ^{2}} ,
\end{align}
can compete with the optical contributions $\delta U _{1}$ and $\delta U _{2}$. For $b \sim \mu$m one finds $\left< \delta V _{\theta 1} \right> \sim Z \times 10 ^{-12}$eV, while the other terms are of the order of $\left< \delta V _{\theta 2} \right> \sim 5 \varepsilon _{1} \times 10 ^{-17}$eV and $\left< \delta V _{\theta 3} \right> \sim  (3 \varepsilon _{1} ^{2} / Z) \times 10 ^{-21}$eV, which are strongly suppressed with respect to $\left< \delta V _{\theta 1} \right>$. Therefore, in the subsequent analysis we only consider the term $\delta V _{\theta 1}$ while disregarding the others. An interesting feature of this term is that the product $\theta \left< L _{z} \right>$ can be positive or negative, depending on both the sign of the magnetization on the surface of the TI and the projection of the $z$ component of the angular momentum. When negative, this term provides a positive contribution for the Hamiltonian, thus in principle  competing with the attractive character of the CP interaction optical contributions in the CP potential. We make a detailed discussion of this possibility in section \ref{Casimir-Polder}. This property is a direct consequence of the TME effect.

The next term to be considered is  $(e ^{2} / 2 m _{e}) \textbf{A} ^{2}$ leading to corrections of the order of
\begin{align}
\frac{\left< (e ^{2} / 2 m _{e}) \textbf{A} ^{2} \right>}{E_g} = - \left[ \frac{ \alpha ^{2} \xi ^{2} (\theta / \pi)(\varepsilon _{1} / 4)}{(\varepsilon _{1} + \varepsilon _{1})(1 / \mu _{1} + 1 / \mu _{2}) + \tilde{\theta} ^{2}} \right] ^{2}. 
\end{align} 
Since $\xi \sim \alpha ^{2}$ for $b \sim \mu$m, this term is $\alpha ^{6} \sim 10 ^{-13}$ smaller than the optical contributions and then will not be taken  into account. Finally we are left with the spin-dependent interaction terms in Eq. (\ref{FULLHAM}), arising from the interaction proportional to  $\textbf{S} \cdot \textbf{B}$. The  most important contribution is 
\begin{align}
\frac{\left< \delta W _{\theta 1} \right>}{E _{g}} = \alpha ^{2} \xi ^{2} \frac{(1-Z) (\theta / \pi) \left< S _{z} \right>}{(\varepsilon _{1} + \varepsilon _{2})(1 / \mu _{1} + 1 / \mu _{1}) + \tilde{\theta} ^{2}} ,
\end{align}
which is of the same order of magnitude than $\left< \delta U _{2} \right>$ and $\left< \delta V _{\theta 1} \right>$ for $b \sim \mu$m and $Z \neq 1$, and vanishes for the hydrogen atom. Therefore, we retain such term in our subsequent analysis. Note that in an analogous fashion to that of the term $\delta V _{\theta 1}$, the sign of this term can be tuned by means of the product $\theta \left< S _{z} \right>$, which can be either positive or negative depending on both the sign of the magnetization on the surface of the TI and $z$ component of the spin. One can further verify that the second term, $\left< \delta W _{\theta 2} \right>$, is smaller than $\left< \delta W _{\theta 1} \right>$ by a factor of $\alpha ^{2} \approx 10 ^{-5}$, and thus it can be discarded.

Finally we make a rough estimation of the weights of the terms not considered in our Hamiltonian (\ref{Hamiltonian}). We first consider the interaction between the image monopole magnetic fields (\ref{MonopolarMagneticFields}) and the nuclear magnetic moment $\boldsymbol{\mu} = (Ze g _{N} / 2 m _{N}) \textbf{I}$, where $\textbf{I}$ is the nuclear spin, $m _{N}$ is its mass and $g _{N}$ is its gyromagnetic ratio. This is given by $\delta Q = - \boldsymbol{\mu} \cdot \textbf{B}$. Performing a Taylor expansion of the magnetic field we find similar expressions to those of Eq. (\ref{DeltaW}). Thus we find that the ratio between the most important contributions, $\delta Q _{1} = Z(1-Z) (eg g _{N} / 8 m _{N} b ^{2}) I _{z}$ and $\delta W _{\theta 1}$, become
\begin{align}
\frac{\left\langle \delta Q _{1} \right\rangle}{\left\langle \delta W _{\theta 1} \right\rangle} = \frac{Z g _{N}}{2} \frac{m _{e}}{m _{N}} . \label{ratioWQ}
\end{align}
As discussed in the previous paragraphs, $\delta W _{\theta 1}$ is of the order of $10 ^{-12}$eV, and thus any smaller contribution can be disregarded in our analysis. Indeed, one can directly verify that the ratio (\ref{ratioWQ}) is very small ($\sim 10 ^{-4}$) and this is why we have not considered the $\delta Q$ interaction in our initial Hamiltonian (\ref{Hamiltonian}). On the other hand, the magnetic moment of the nucleus $\boldsymbol{\mu}$ will induce an image electric dipole $\textbf{d}$ due to the TME, whose electric field $\textbf{E}$ will in turn interact with the atomic dipole moment $\textbf{p}$ according to $\delta P = - \textbf{p} \cdot \textbf{E}$. In this case the full expression is rather complicated but a rough estimation can be done. We can naively think that the magnetic dipole moment is sourced by an elementary electric current $\textbf{j}$ whose magnitude must be proportional to $\vert \boldsymbol{\mu} \vert$. This implies that the interaction $\delta P$ must be proportional to $\vert \boldsymbol{\mu} \vert$ (from the source) and to $g$ (from the nondiagonal components of the Green's function). These simple arguments imply that $\left\langle \delta P \right\rangle / \left\langle \delta V _{\theta 1} \right\rangle \approx (g _{N} / 2) ( m _{e} / m _{N})$, which is small enough to be considered in our analyses.

The previous order of magnitude estimations leave us with
\begin{equation}
H = H _{0} + \delta U _{1} + \delta U _{2} + \delta V _{\theta 1} + \delta W _{\theta 1} , \label{ReducedHamiltonian}    
\end{equation}
as the final Hamiltonian describing the ion-TI interaction, to be considered in the next sections. Here, each term is given by 
\begin{align}
\delta U _{1} &= \frac{\varepsilon _{1}}{2} \frac{1 - Z}{Z} \kappa \xi ^{2} E _{g} \tilde{z} , \label{U1} \\ \delta U _{2} &= - \frac{\varepsilon _{1} ^{2}}{8 Z ^{2}} \kappa \xi ^{3} E _{g} \left[ Z \tilde{r} ^{2} - (3 Z - 4) \tilde{z} ^{2} \right] , \label{U2} \\ \delta V _{\theta 1} &= \frac{Z}{4} ge \xi ^{2} E _{g} L _{z} , \label{V1} \\ \delta W _{\theta 1} &= \frac{1 - Z}{2}  ge \xi ^{2} E _{g} S _{z} , \label{W1}
\end{align}
where we have defined the escaled coordinates $\tilde{x} _{i} \equiv (Z / \varepsilon _{1}) (x _{i} / a _{0})$.

According to the statement of the problem, we are considering a hydrogenlike ion embedded in a medium with optical properties $(\varepsilon _{1} , \mu _{1})$ at a distance $b$ from a planar topological insulator charaterized by its optical properties $(\varepsilon _{2} , \mu _{2})$ and the TMEP $\theta$, as shown in Fig. \ref{figure}. Without loss of generality we can restrict our analysis to the case $\mu _{1} = \mu _{2} = 1$, which is suitable for both conventional and topological insulators. Furthermore, we observe that the potentials $\delta V _{\theta 1}$ and $\delta W _{\theta 1}$ are exclusively of topological origin in the sense that they vanish for $\theta = 0$. On the contrary, the optical and topological properties coexist for the potentials $\delta U _{1}$ and $\delta U _{2}$ provided $\varepsilon _{1} \neq \varepsilon _{2}$ since they depend on $\kappa$, defined in Eq. (\ref{kappa}). Therefore, one can consider the following two interesting cases: a) $\varepsilon _{1} = \varepsilon _{2}$ and b) $\varepsilon _{1} \neq \varepsilon _{2}$. In the former case, we consider the ion to be embedded in a dielectric medium with the same optical properties  that those of the TI, such that the electrostatic effects are suppressed and only the topological ones become important. The second case is perhaps the most realistic situation from the experimental point of view since spectroscopy experiments consider the atoms in  vacuum. On the other hand, from the potentials (\ref{U1})-(\ref{W1}) we can also distinguish two situations of interest, i.e. i) $Z \neq 1$ (hydrogenlike ions) and ii) $Z = 1$ (hydrogen atom). The fundamental difference arises from the fact that potentials $\delta U _{1}$ and $\delta W _{\theta 1}$ vanish for the hydrogen atom case. In the next sections we discuss the lowest lying energy levels, in each case, where the TME effects become manifest.

\section{Energy shifts of the spectrum of hydrogenlike ions}
\label{Energy-Levels}

\subsection{General considerations}

In this section we work out the energy shifts on the hyperfine structure  states of hydrogenlike ions due to the Casimir-Polder interaction $\delta U _{1} + \delta U _{2} +\delta V _{\theta 1} + \delta W _{\theta 1}$. We study the cases described in the end of the previous section but we left the case of the hydrogen atom in vacuum for a detailed analysis in the next section.

In our notation, the electron variables are labeled by the quantum numbers $n$, $\ell$, $s$, $j$ and $m _{j}$, where the total electron angular momentum $\mathbf{J} = \mathbf{L} + \mathbf{S}$ is labeled by $j = \ell \pm s$, with $\ell$ being the orbital angular momentum quantum number and $s = 1/2$ its spin. Explicit forms for the fine structure states, abbreviated as $\left| j, m _{j} \right\rangle _{\mbox{\scriptsize fs}} \equiv \left| n, \ell ,s,j,m_{j}\right\rangle$, are 
\begin{align}
\left| j,m_{j}\right\rangle _{\mbox{\scriptsize fs}} &=\sum_{m_{s}=-s}^{+s}\left\langle \ell ,s,m_{j}-m_{s},m_{s}|j,m_{j}\right\rangle \times  \notag \\ & \left| n,\ell ,m_{j}-m_{s}\right\rangle \otimes \left| s,m_{s}\right\rangle _{e},  \label{FineineBasis}
\end{align}
where the $\left| n,\ell ,m\right\rangle $ are the spinless Coulomb bound states with $\left\langle \mathbf{r}|n,\ell ,m\right\rangle = R _{n \ell
}(r)Y_{\ell }^{m}(\vartheta ,\varphi)$, the $\left| s , m _{s} \right\rangle _{e}$ are the electron spin states. The required Clebsch-Gordan (CG)
coefficients are given by 
\begin{align}
\left\langle \ell ,s,m_{j}\mp s,\pm s|j=\ell +s,m_{j}\right\rangle & = \phantom{\mp}\sqrt{\frac{1}{2}\pm \frac{m_{j}}{2\ell +1}},  \notag \\ \left\langle \ell ,s,m_{j}\mp s,\pm s|j=\ell -s,m_{j}\right\rangle & =\mp  \sqrt{\frac{1}{2}\mp \frac{m_{j}}{2\ell +1}}.  \label{Clebsch-Gordan}
\end{align}
The states $\left| n, \ell , s , j , m _{j} \right\rangle $ are, by construction, simultaneous eigenfunctions of $\mathbf{L} ^{2}$, $\mathbf{S} ^{2}$, $\mathbf{J} ^{2}$ and $J _{z}$.

At the hyperfine level we must include the nuclear spin $i$. The total atomic angular momentum $\mathbf{F} = \mathbf{J} + \mathbf{I}$ has quantum number $f$ satisfying $\vert j - i \vert \leq f \leq \vert j + i \vert$. It is conserved due to rotational symmetry, so the states having different eigenvalues $m _{f}$ of $F _{z}$ would be degenerate in the absence of an external magnetic field, but all other degeneracies are broken. The hyperfine structure states, abbreviated as $\left| j,f,m _{f} \right\rangle _{\mbox{\scriptsize hfs}}\equiv \left| n,\ell , s , j , f , m _{f} \right\rangle$, have the form 
\begin{align}
\left| j,f,m _{f}\right\rangle _{\mbox{\scriptsize hfs}} &= \sum _{m _{i} = - i} ^{+i} \left\langle j,i,m _{f} - m _{i} , m _{i} \vert f,m_{f}\right\rangle \times  \notag \\ & \left| j,m _{f} - m _{s} \right\rangle _{\mbox{\scriptsize fs}} \otimes \left| i,m _{i} \right\rangle _{N},  \label{HyperfineineBasis}
\end{align}
where $\left| i,m _{i} \right\rangle _{N}$ are the nuclear spin states. The CG coefficients in Eq. (\ref{HyperfineineBasis}) depend on the value of the nuclear spin. For hydrogenlike ions with $i = 1/2$ we have $f = j \pm ^{\prime} i$, and the CG coefficients are given by (\ref{Clebsch-Gordan}) with the replacement $\left\{ s,j,l,m_{j}\right\} \rightarrow \left\{i,f,j,m _{f}\right\} $. For spin $i > 1/2$ the expressions for the CG coefficients are simple but more cumbersome than those appearing in (\ref{Clebsch-Gordan}). The radial contribution of the hyperfine states, which we take as the radial functions $R _{n \ell}$ of the Coulomb potential $-Ze^{2}/\varepsilon _{1}r$, are the zeroth order aproximation of the full eigenfunctions in the Hamiltonian including the fine and hyperfine structure contributions, with their first order correction being of the order $\alpha ^{2}$. Since all terms in the potentials (\ref{U1})-(\ref{W1}) are already of higher order, this approximation is enough to compute the lowest order additional energy shifts.

We are start by perturbing the hyperfine atomic spectrum, which is nondegenerate except for the quantum number $m _{f}$, so that the potential $\delta U _{1}$ do not contribute to the first order energy shifts since it is a first rank spherical tensor. Nevertheless it contributes to second order shifts, but its order of magnitude will be suppressed by a factor of $\alpha ^{2} \sim 10 ^{-5}$ with respect to the other potentials (\ref{U2})-(\ref{W1}) for $b \sim \mu$m. Thus, in the following we do not consider such term. In a similar fashion, although the potential $\delta U _{2}$ contributes both to first and second order energy shifts, it is sufficient to consider only the former contribution. The perturbation $\delta U _{2}$ does not depend on the nuclear spin and its expectation value can be directly computed in the hyperfine structure basis. The result is 
\begin{equation}
\frac{\left\langle \delta U_{2}\right\rangle }{E_{g}}=-\frac{\kappa \xi ^{3}%
}{8}\left( \frac{\varepsilon _{1}}{Z}\right) ^{2}\left\langle \tilde{r}%
^{2}\right\rangle _{n\ell }\Sigma _{Z}^{ijfm_{f}},  \label{DELTAU2}
\end{equation}
where 
\begin{align}
\left\langle \tilde{r}^{2}\right\rangle _{n\ell }& =\frac{1}{2}n^{2}\left[
5n^{2}-3\ell (\ell +1)+1\right] ,  \label{R2} \\
\Sigma _{Z}^{ijfm_{f}}& =Z-(3Z-4)\left\langle \cos ^{2}\vartheta
\right\rangle _{\mbox{\scriptsize hfs}}^{ijfm_{f}}.  \label{Sigma}
\end{align}
The expectation value in the hyperfine structure basis can be computed in terms of those in the fine structure basis as 
\begin{align}
\left\langle \cos ^{2}\vartheta \right\rangle _{\mbox{\scriptsize hfs}}^{ijfm_{f}}& =\sum_{m_{i}=-i}^{+i}\left\langle j,i,m_{f}-m_{i},m_{i}|f,m_{f}\right\rangle ^{2}\times  \notag \\ & \hspace{2cm}\left\langle \cos ^{2}\vartheta \right\rangle _{\mbox{\scriptsize fs}}^{j,m_{f}-m_{i}},  \label{ExpectValue}
\end{align}
where 
\begin{equation}
\left\langle \cos ^{2}\vartheta \right\rangle _{\mbox{\scriptsize fs}}^{j,m_{j}}=\frac{j(j+1)-m _{j}^{2}}{2j(j+1)}.
\end{equation}
The final form of Eq. (\ref{ExpectValue}) strongly depends on the value of the nuclear spin. For example, for $i=1/2$ we have $f = j \pm ^{\prime} i$, and a simple calculation yields 
\begin{equation}
\left\langle \cos ^{2}\vartheta \right\rangle _{\mbox{\scriptsize hfs}} ^{ijfm_{f}}=\frac{1}{2}-\frac{1+4m_{f}^{2}}{8j(j+1)}\pm ^{\prime }\frac{m _{f}^{2}}{j(j+1)(2j+1)}.
\end{equation}
Analogous expressions for nuclear spin $i>1/2$ can be obtained in a similar manner.

The energy shifts arising from the Zeeman-like potentials (\ref{V1}) and (\ref{W1}) can be computed in a simple fashion. We consider the potential 
\begin{equation}
\delta V _{\theta} \equiv \delta V _{\theta 1} +\delta W _{\theta 1} = \frac{Z}{4} \, g e \xi ^{2}  E _{g} V _{z},  \label{Ea}
\end{equation}
where we have defined the operator 
\begin{equation}
\mathbf{V} = \mathbf{L} - 2 \frac{Z-1}{Z}\mathbf{S} .  \label{A-Operatr}
\end{equation}
We observe that the perturbation potential $\delta V _{\theta}$ is of topological origin, since it vanishes for $\theta = 0$, and therefore it is a signature of the topological nontriviality of the TIs and particularly of the image magnetic monopole effect. We are in a subspace of fixed $n$, $\ell$, $j$ and $s$, so we can use the Wigner-Eckart theorem to make the replacement $V _{z}= g _{\mbox{\scriptsize fs}} ^{(Z)} J _{z}$, provided we never compute matrix elements between states with different $j$. Here $g _{\mbox{\scriptsize fs}} ^{(Z)} = \left\langle \mathbf{J} \cdot \mathbf{V} \right\rangle /j(j+1)$ is a fine-structure type $g$-factor. Since we are also in a subspace with fixed $f$, we can use Wigner-Eckart theorem again to take $J _{z} = g _{\mbox{\scriptsize hfs}} F _{z}$, where $g _{\mbox{\scriptsize hfs}} = \left\langle \mathbf{J}\cdot \mathbf{F} \right\rangle / f(f+1)$ is the usual hyperfine-structure $g$-factor. Therefore, the perturbation $\delta V _{\theta}$ lifts the degeneracy between hyperfine levels with equal $f$ and unequal $m _{f}$ in a linear fashion, giving energy shifts 
\begin{equation}
\left\langle \delta V _{\theta} \right\rangle =\frac{Z}{4} ge \xi ^{2} E _{g} g _{\mbox{\scriptsize fs}} ^{(Z)} g _{\mbox{\scriptsize hfs}} m _{f} ,  \label{EnergyShiftsA}
\end{equation}
where the $g$-factors are given by 
\begin{align}
g _{\mbox{\scriptsize fs}} ^{(Z)} &= \frac{1}{2Z}\left[ (2-Z) + (3Z-2) \frac{\ell (\ell + 1) - s (s+1)}{j(j+1)}\right] ,  \notag \\ g _{\mbox{\scriptsize hfs}} &= \frac{1}{2}\left[ 1+\frac{j(j+1)-i(i+1)}{f(f+1)} \right] .  \label{g-factors}
\end{align}
Now let us discuss the conditions under which the energy shifts (\ref{EnergyShiftsA}) induced by the TME are comparable with the nonperturbed hyperfine spectrum for different cases. To this end, we consider the expression for the hyperfine splitting of a one-electron ion \cite{Shabaev}
\begin{equation}
E _{\mbox{\scriptsize hfs}} = \frac{\alpha \left( \alpha Z\right) ^{3}}{\varepsilon ^{3}n^{3}}g_{\mu }\frac{m_{e}^{2}}{m_{p}}\frac{f(f+1)-i(i+1)-j(j+1)}{2j(j+1)(2\ell +1)},  \label{HyperfineHIon}
\end{equation}
where $m_{p}$ is the proton mass, $g_{\mu }=\mu /(\mu _{N}i)$, $\mu $ is the nuclear magnetic moment and $\mu _{N}$ is the nuclear magneton. In the above we have neglected the relativistic corrections together with the nuclear charge distribution correction, the Bohr-Weisskopf correction and the radiative corrections. Also we have incorporated the effect of the dielectric medium with permitivity $\varepsilon$. 

From equations (\ref{DELTAU2}) and (\ref{EnergyShiftsA}) we find two different regimes to be analyzed separately. On the one hand, we observe that for nucleus-surface distances of the order of micrometers ($b \sim \mu$m), the topological contributions are suppressed with respect to the standard electromagnetic ones ($\theta = 0$) provided $\varepsilon _{1} \neq \varepsilon _{2}$. On the other hand we can see that the case $\varepsilon _{1} = \varepsilon _{2}$ enhance the topological contribution (\ref{EnergyShiftsA}), and thus it deserves a separate analysis. Next, we analyze the cases mentioned above. Let us consider both: a) the lowest and b) the highest lines where the TME becomes manifest, that is, the ground state $ 1$S$_ {1/2}$ and the circular Rydberg states, respectively. For definiteness, in the sequel we restrict our analysis to the recently discovered topological insulator TlBiSe$_{2}$ for which $\varepsilon _{2} = 4$ and $\mu _{2} = 1$, and we left $\theta$ as a free parameter.

\subsection{Case $\varepsilon _{1} = \varepsilon _{2}$} \label{Case1}

When the hydrogenlike ion is embedded in a medium with the same optical properties of the TI, i.e. $\varepsilon _{1} = \varepsilon _{2} \equiv \varepsilon$, the energy shifts are given by Eqs. (\ref{DELTAU2}) and (\ref{EnergyShiftsA}) together with
\begin{equation}
\kappa = - \frac{\tilde{\theta} ^{2}}{4 \varepsilon ^{2}} \quad , \quad e g =\frac{\alpha \tilde{\theta}}{2\varepsilon} , \label{EP1EP2}
\end{equation}
where we have considered that $\varepsilon \gg \alpha ^{2}$. 

\subsubsection{The ground state $1 \mbox{S} _{1/2}$}

The hyperfine spectrum for the ground state $1 \mbox{S} _{1/2}$ is: 
\begin{equation}
E _{\mbox{\scriptsize hfs}} ^{1 \mbox{\scriptsize S} _{1/2}} =\frac{Z^{3}}{\varepsilon ^{3}}\Lambda g _{\mu} \left[ f ( f + 1 ) - i ( i + 1 ) - 3/4 \right] ,  \label{HyperfineHIon1S1}
\end{equation}
where $\Lambda \equiv (4/3) \vert E _{g} \vert \alpha ^{2} (m _{e} / m _{p} ) = 5.25 \times 10 ^{-7}\,$eV$= 1.27 \times 10 ^{8}\,$Hz. One can further check that Eq. (\ref{HyperfineHIon1S1}) correctly yields the $21$ cm line arising from the transition between the states with $f =1$ and $f =0$ in hydrogen, for which $i = 1/2$, $g _{\mu} = 5.56$ and $Z = 1$. On the other hand, from Eq. (\ref{EnergyShiftsA}) together with Eq. (\ref{EP1EP2}) the topological contribution gives \begin{equation}
\langle \delta V _{\theta} \rangle _{1 \mbox{\scriptsize S} _{1/2}} = \Gamma \frac{Z - 1}{\varepsilon} (\theta / \pi ) m _{f} \left[ \frac{f (f + 1) - i (i + 1) + 3/4}{f ( f + 1 )} \right] ,
\end{equation}
where $\Gamma \equiv \vert E _{g} \vert \alpha ^{2} \xi ^{2} / 8 = 2.57 \times 10 ^{-13}\,$eV$= 62.2\,$Hz for $b \sim \mu$m. The ratio between the hypefine spectrum and the topological contribution for an hydrogenlike ion for which $i = 1/2$, $f = 1$ and $m _{f} = \pm 1$, is
\begin{align}
r _{\theta \mbox{\scriptsize -hfs}} ^{1 \mbox{\scriptsize S} _{1/2}} = \frac{\vert \langle \delta V _{\theta} \rangle _{1 \mbox{\scriptsize S} _{1/2}} \vert}{\vert E _{\mbox{\scriptsize hfs}} ^{1 \mbox{\scriptsize S} _{1/2}} \vert} = \frac{\vert \theta / \pi \vert}{g _{\mu}} \frac{Z-1}{Z ^{3}} \left( 1.5 \times 10 ^{-5} \right) . \label{ratio}
\end{align}
One can further check that $r _{\theta \mbox{\scriptsize -hfs}} ^{1 \mbox{\scriptsize S} _{1/2}}$ has a maximum at $Z = 3/2$ and decreases as increasing $Z$, thus impliying that hydrogenlike ions with small values $Z$ are the best probes to test the TME in its ground state. Consider, for example, the $^{3}$He$^{+}$ ion, for which $Z =2$, $i = 1/2$ and $\mu / \mu _{N} = 1.15$. In this case, the ratio (\ref{ratio}) becomes $r _{\theta \mbox{\scriptsize -hfs}} ^{1 \mbox{\scriptsize S} _{1/2}} = \vert \theta / \pi \vert (3.4 \times 10 ^{- 6})$, which is small enough to be measured for appropriate values of the TMEP. For heavy ions, such as $^{207}$Pb$^{81+}$, for which $Z=82$, $i = 1/2$ and $\mu / \mu _{N} = 0.587$, we find $r _{\theta \mbox{\scriptsize -hfs}} ^{1 \mbox{\scriptsize S} _{1/2}} = \vert \theta / \pi \vert (7.5 \times 10 ^{- 9})$, which is even smaller than those for the $^{3}$He$^{+}$ ion.

\subsubsection{Circular Rydberg states}

Now let us consider the case of circular Rydberg hidrogenlike ions, i.e. highly excited states with its quantum numbers maximally projected. We define the circular states as $\left| n \right> _{\mbox{\scriptsize circ}} = \left| n , \ell _{\mbox{\scriptsize max}} = n - 1 , j _{\mbox{\scriptsize max}} = \ell _{\mbox{\scriptsize max}} + 1/2 , f _{\mbox{\scriptsize max}} = j _{\mbox{\scriptsize max}} + i , \right. $ $\left. m _{f \, \mbox{\scriptsize max}} = f _{\mbox{\scriptsize max}} \right>$. Adapting the approach of Ref. \cite{Barton} for Rydberg hydrogen to our case we obtain a retardation line given by 
\begin{align}
\vert \delta E _{r} \vert &= \frac{Z^{2}}{\varepsilon ^{2} n ^{3}} \, \left( 6.58 \times 10 ^{15} \, \mbox{Hz} \right) \\ L _{r} &= \frac{\varepsilon ^{2} n ^{3}}{Z ^{2}} \, \left( 4.56 \times 10 ^{-2} \, \mu \mbox{m} \right) .  \label{RETLINE}
\end{align}
According to Eq. (\ref{HyperfineHIon}) the energy difference $\Delta E _{\mbox{\scriptsize hfs}} ^{\mbox{\scriptsize circ}}$ between neighboring hyperfine circular Rydberg states $\left| n \right> _{\mbox{\scriptsize circ}}$ and $\left| n - 1 \right> _{\mbox{\scriptsize circ}}$ is
\begin{equation}
\Delta E _{\mbox{\scriptsize hfs}} ^{\mbox{\scriptsize circ}} = \frac{5 i g _{\mu}}{n ^{6}} \frac{Z ^{3}}{\varepsilon ^{3}} \, \left( 4.76 \times 10 ^{8} \, \mbox{Hz} \right) .  \label{DELTAHFS}
\end{equation}
Our main concern is with the energy shifts $\delta U _{2}$ and $\delta V _{\theta}$ given by Eqs. (\ref{DELTAU2}) and (\ref{EnergyShiftsA}), respectively. In our approximation, which is that of circular Rydberg states embedded in a medium with the same optical properties to that of the TI, we find that $g _{\mbox{\scriptsize hfs}} = g _{\mbox{\scriptsize fs}} ^{(Z)} = 1$, from which we establish the following ratio
\begin{align}
r _{\theta \mbox{\scriptsize -hfs}} ^{\mbox{\scriptsize circ}} &\equiv \frac{\vert \left\langle \delta V _{\theta} \right\rangle \vert}{\Delta E _{\mbox{\scriptsize hfs}} ^{\mbox{\scriptsize circ}}} = \frac{(\theta / \pi) \varepsilon ^{2} \xi ^{2} n ^{7}}{40 Z ^{2} i g _{\mu}} \frac{m _{p}}{m _{e}} .  \label{r1}
\end{align}
In a similar fashion, we can also establish an expression for the ratio between the maximum energy shift $\vert \left\langle \delta U _{2} \right\rangle _{\mbox{\scriptsize max}} \vert$, which is obtained from Eq. (\ref{DELTAU2}) together with $\left\langle \cos ^{2} \vartheta \right\rangle = 1$ and $\left\langle \tilde{r} ^{2} \right\rangle = n ^{4}$, and the hyperfine energy $\Delta E _{\mbox{\scriptsize hfs}} ^{\mbox{\scriptsize circ}}$. We obtain
\begin{align}
r _{U \mbox{\scriptsize -hfs}} ^{\mbox{\scriptsize circ}} &\equiv \frac{\vert \left\langle \delta U _{2} \right\rangle _{\mbox{\scriptsize max}} \vert}{\Delta E _{\mbox{\scriptsize hfs}} ^{\mbox{\scriptsize circ}}} = \frac{(\theta / \pi) ^{2} \varepsilon ^{3} \xi ^{3} n ^{10}}{80 Z ^{4} i g _{\mu}} \frac{m _{p}}{m _{e}} . \label{r2}
\end{align}
We observe that high values of the TMEP $\theta$ favor the ratios (\ref{r1}) and (\ref{r2}), therefore we take $\theta = 11 \pi$ hereafter. Using the numerical values $\varepsilon = 4$ and $m _{p} / m _{e} = 1840$ we find that the ratios become
\begin{align}
r _{\theta \mbox{\scriptsize -hfs}} ^{\mbox{\scriptsize circ}} = \frac{\xi ^{2} n ^{7}}{i g _{\mu} Z ^{2}} \, \left( 8 \times 10 ^{3} \right) \, , \, r _{U \mbox{\scriptsize -hfs}} ^{\mbox{\scriptsize circ}} = \frac{\xi ^{3} n ^{10}}{i g _{\mu} Z ^{4}} \, \left( 3.6 \times 10 ^{5} \right) . \label{r1-2}
\end{align}
\begin{figure}[tbp]
\begin{center}
\includegraphics[scale=0.38]{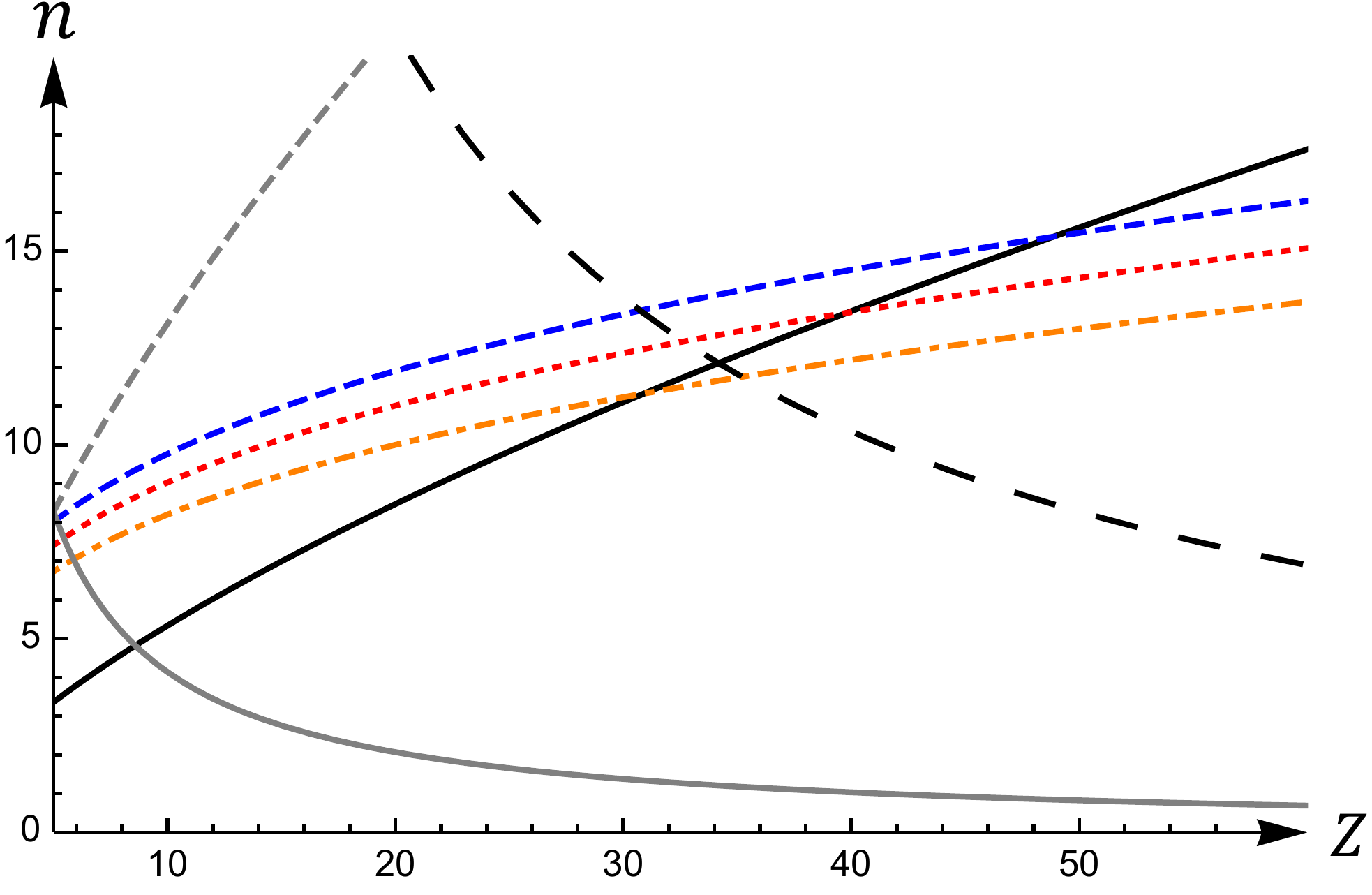}
\end{center}
\caption{{\protect\small Parameter region $n (Z)$, for $b = B _{\mbox{\scriptsize min}}$, for detecting the TME in circular Rydberg hydrogenlike ions for $\varepsilon _{1} = \varepsilon _{2}$. The meaning of each line is given in the text. }}
\label{paramaterregionb}
\end{figure}
We now come to the problem of choosing an adequate value for $b$. The lowest value of $b$ which will make the ratios (\ref{r1-2}) as high as possible is limited by the thickness $w \approx 6 \times 10 ^{-3} \, \mu$m ($\xi \approx 10 ^{-2}$) of the magnetic coating, but more importantly by the experimental possibilities. Motivated by the works in Refs. \cite{Haroche1, Haroche2} we take $b \leq \mu$m and explore the range $ B_{\mbox{\scriptsize min}} < b < B_{\mbox{\scriptsize max}}$, where $B_{\mbox{\scriptsize min}} = 0.265 \mu$m and $B _{\mbox{\scriptsize max}} = 1.1 \mu$m. Notice that for $b = B _{\mbox{\scriptsize min}}$, we have $b = 44.2 \, w$ which is still larger than the width of the magnetic coating. Now let us discuss the allowed parameter region for $b =B_{\mbox{\scriptsize min}}$.

By imposing the nonretarded constraint $B_{\mbox{\scriptsize max}} \leq L _{r}$ we obtain the condition $n \geq 1.15 \ Z ^{2/3}$, which corresponds to the continuous black line in Fig. \ref{paramaterregionb}. Since the energy shifts must be smaller than the unperturbed energy spectrum we must consider that $r _{\theta \mbox{\scriptsize -hfs}} ^{\mbox{\scriptsize circ}} < 1$, thus providing the condition $n \leq 3.97 \left( i g _{\mu} Z ^{2} \right) ^{1/7}$ for $b = B_{\mbox{\scriptsize min}}$. In Fig. \ref{paramaterregionb} we show the latter condition for $i g _{\mu} = 5.53$ (blue dashed line), $i g _{\mu} = 3.18$ (red dotted line) and $i g _{\mu} = 1.62$ (orange dot-dashed line), which fall  within the nonretarded region for $Z \leq 49, \, 39, \, 30$, respectively. We have chosen the values for $i g _{\mu}$ from Ref. \cite{Shabaev}. In Fig. \ref{paramaterregionb} we also show the curves $n (Z)$ corresponding to the lower limits of the regions satisfying $\vert \left\langle \delta V _{\theta} \right\rangle \vert > 10 ^{6}$Hz (black large dashed line) and $\vert \left\langle \delta V _{\theta} \right\rangle \vert > 10 ^{5}$Hz (continuous gray line) for $b = B _{\mbox{\scriptsize min}}$. One can further verify that the condition $\vert \left\langle \delta V _{\theta} \right\rangle \vert > 10 ^{6}$Hz for $b = B_{\mbox{\scriptsize max}}$ selects values of $n$ larger than $100$ for the whole interval $1 < Z < 100$, and this is why we have restricted ourselves to $b = B_{\mbox{\scriptsize min}}$ as a reasonable lower limit for $b$. The region below the gray dashed line corresponds to the condition $r _{\theta \mbox{\scriptsize -hfs}} ^{\mbox{\scriptsize circ}} > 10 \, r _{U \mbox{\scriptsize -hfs}} ^{\mbox{\scriptsize circ}}$ for $b = B_{\mbox{\scriptsize min}}$; while the case $r _{\theta \mbox{\scriptsize -hfs}} ^{\mbox{\scriptsize circ}} > r _{U \mbox{\scriptsize -hfs}} ^{\mbox{\scriptsize circ}}$ is not shown in the figure since it lies higher than the gray dashed line. For the ions listed in Table 3 of Ref. \cite{Shabaev} we find that the maximum value for the TME correction is $\vert \left\langle \delta V _{\theta} \right\rangle \vert = 1.83 \times 10 ^{6}$Hz for $^{113}$In$^{48+}$.

To close this section we  remark that the topological Zeeman-type energy shifts can be of the same order of magnitud to that of the hyperfine energy levels of a hydrogenlike atom embedded in a medium with the same dielectric constant to that of the TI. From Eq. (\ref{r1}) we have
\begin{equation}
n=5.59 \times\sqrt[7]{ \left( \frac{\mu }{\mu _{N}}\right) Z^{2}}
\label{QuantumNumber}
\end{equation}
for $r _{\theta \mbox{\scriptsize -hfs}} ^{\mbox{\scriptsize circ}} = 1$. In Table \ref{TableIons} we present the values of the principal quantum number which solves Eq. (\ref{QuantumNumber}) for different circular states of hydrogenlike heavy ions.

\begin{table}[th]
\centering                                          
\begin{tabular}{ccccc}
\hline\hline
Ion & $Z$ & $\mu / \mu _{N}$ & $i$ & $n$ \\[0.5ex] \hline
${}^{133}\mbox{Cs} ^{54+}$ & 55 & 2.5825 & 7/2 & 20 \\ 
${}^{159}\mbox{Tb} ^{64+}$ & 65 & 2.014 & 3/2 & 20 \\ 
${}^{207}\mbox{Pb} ^{81+}$ & 82 & 0.587 & 1/2 & 18 \\ 
${}^{235}\mbox{U} ^{91+}$ & 92 & 0.39 & 7/2 & 18 \\[1ex] \hline
\end{tabular}
\caption{The quantum number for different heavy hydrogenlike ions satisfying $r _{\theta \mbox{\scriptsize -hfs}} ^{\mbox{\scriptsize circ}} = 1$.}
\label{TableIons}
\end{table}

\subsection{Case $\varepsilon _{1} \neq \varepsilon _{2}$} \label{Case2}
 
Now let us consider the atom to be embedded in a medium with different optical propeties to that of the TI, i.e. $\varepsilon _{1} \neq \varepsilon _{2}$. For definitness here we consider the atom in  vacuum, such that the basic parameters now become
\begin{equation}
\kappa = \frac{1 - \varepsilon _{2}}{1 + \varepsilon _{2}} \quad , \quad eg = \frac{\alpha ^{2} (\theta / \pi)}{1 + \varepsilon _{2}} , \label{EPINEQEP2}
\end{equation}
where we used that $\tilde{\theta} ^{2} \ll 1$. We observe that $\vert eg / \kappa \vert \sim 2 \times 10 ^{-4}$ for the TI TlBiSe$_{2}$; therefore, contrary to the previous situation in section \ref{Case1}, the topological correction $\langle \delta V _{\theta }\rangle$ will be much supressed with respect to the optical correction $\left\langle \delta U _{2}\right\rangle$.

\subsubsection{The ground state $1 \mbox{S} _{1/2}$}

Using the hyperfine structure energy levels together with the Zeeman-type energy shifts for the ground state $1$S$_{1/2}$, we find the ratio
\begin{align}
t _{\theta \mbox{\scriptsize -hfs}} ^{1 \mbox{\scriptsize S} _{1/2}} = \frac{\vert \langle \delta V _{\theta} \rangle _{1 \mbox{\scriptsize S} _{1/2}} \vert}{\vert E _{\mbox{\scriptsize hfs}} ^{1 \mbox{\scriptsize S} _{1/2}} \vert} = \frac{\vert \theta / \pi \vert}{g _{\mu}} \frac{Z-1}{Z ^{3}} \left( 8 \times 10 ^{-8} \right) , \label{ratioCase2}
\end{align}
which is three orders of magnitude smaller than those of Eq. (\ref{ratio}). Therefore, the ground state of a hydrogenlike ion in the vacuum is not a good probe to test the TME.

\subsubsection{Circular Rydberg states}

\begin{figure}[tbp]
\begin{center}
\includegraphics[scale=0.38]{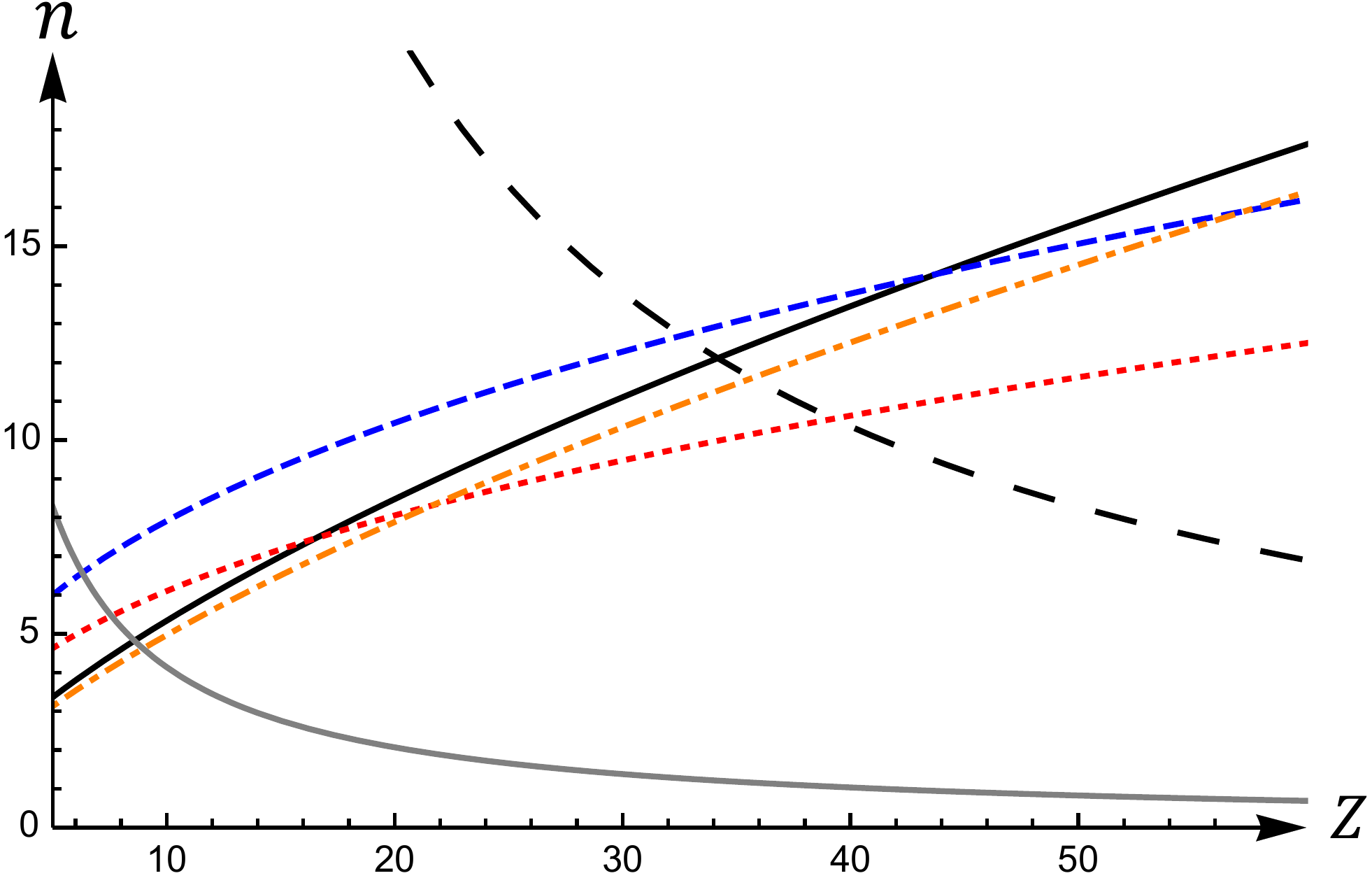}
\end{center}
\caption{{\protect\small Parameter region $n (Z)$, for $b = B _{\mbox{\scriptsize min}}$, for detecting the TME in circular Rydberg hydrogenlike ions for $\varepsilon _{1} \neq \varepsilon _{2}$. The meaning of each line is given in the text. }}
\label{paramaterregionb2}
\end{figure}

The energy difference $\Delta E _{\mbox{\scriptsize hfs}} ^{\mbox{\scriptsize circ}}$ between neighboring hyperfine circular Rydberg states $\left| n \right> _{\mbox{\scriptsize circ}}$ and $\left| n - 1 \right> _{\mbox{\scriptsize circ}}$ is given by Eq. (\ref{DELTAHFS}) with $\varepsilon = 1$. On the other hand, in the approximation we are working with, together with the choice of the parameters (\ref{EPINEQEP2}) of this case, we find the energy shifts $\left\langle \delta U _{2} \right\rangle$ and $\left\langle \delta V _{\theta} \right\rangle$ to be
\begin{align}
\vert \left\langle \delta U _{2} \right\rangle \vert = 1974 \frac{n ^{4}}{Z} \, \mbox{Hz} \quad , \quad \vert \left\langle \delta V _{\theta} \right\rangle \vert = 2408 \, n Z \, \mbox{Hz} , \label{E-shifts}
\end{align}
where we have used that $g _{\mbox{\scriptsize hfs}} = g _{\mbox{\scriptsize fs}} ^{(Z)} = 1$, $\left\langle \tilde{r} ^{2} \right\rangle = n ^{4}$ and $\theta = 11 \pi$. The ratios between the energy shifts (\ref{E-shifts}) and the hyperfine energy difference $\Delta E _{\mbox{\scriptsize hfs}} ^{\mbox{\scriptsize circ}}$ read
\begin{align}
t _{\theta \mbox{\scriptsize -hfs}} ^{\mbox{\scriptsize circ}} = \frac{n ^{7}}{i g _{\mu} Z ^{2}} \, \left( 1 \times 10 ^{-6} \right) \, , \, t _{U \mbox{\scriptsize -hfs}} ^{\mbox{\scriptsize circ}} = \frac{n ^{10}}{i g _{\mu} Z ^{4}} \, \left( 8.3 \times 10 ^{-7} \right) , \label{r-Case2}
\end{align}
for $b = B_{\mbox{\scriptsize min}} = 0.265 \, \mu$m. Now let us analyze the parameter region $n$-$Z$. We first recall that the nonretarded constraint $B_{\mbox{\scriptsize min}} < L _{r}$ provides the condition $n > 1.15 Z ^{2/3}$, which corresponds to the continuous black line in Fig. \ref{paramaterregionb2}.

By imposing the energy shifts to be smaller than the unperturbed energy spectrum, i.e. $t _{\theta \mbox{\scriptsize -hfs}} ^{\mbox{\scriptsize circ}} < 1$, we find the condition $n \leq 2.67 \left( i g _{\mu} Z ^{4} \right) ^{1/10}$ for $b = B _{\mbox{\scriptsize min}}$. Using the Table 3 in Ref. \cite{Shabaev} for the properties of different hydrogenic ions, in Fig. \ref{paramaterregionb2} we show the latter condition for the largest (blue dashed line) and the lowest (red dotted line) values of $ig _{\mu}$, respectively. On the other hand, the condition $t _{\theta \mbox{\scriptsize -hfs}} ^{\mbox{\scriptsize circ}} \leq t _{U \mbox{\scriptsize -hfs}} ^{\mbox{\scriptsize circ}}$ produces the region $n \geq 1.07 Z ^{2/3}$, which corresponds to the orange dot-dashed line in Fig. \ref{paramaterregionb2}.

Fig. \ref{paramaterregionb2} shows that the condition $t _{\theta \mbox{\scriptsize -hfs}} ^{\mbox{\scriptsize circ}} = t _{U \mbox{\scriptsize -hfs}} ^{\mbox{\scriptsize circ}}$ is below  to the retardation line (continuous black line), in such a way that here we always have $t _{\theta \mbox{\scriptsize -hfs}} ^{\mbox{\scriptsize circ}} < t _{U \mbox{\scriptsize -hfs}} ^{\mbox{\scriptsize circ}}$. The region between the large black dashed and the continuous gray lines corresponds to $10 ^{6}$Hz $< \vert \langle \delta V _{\theta} \rangle \vert < 10 ^{7}$Hz. Therefore, we observe that the condition $\varepsilon
_{1} \neq \varepsilon _{2}$ places a strong restriction upon the allowed parameter region, when compared with the similar situation in the case $\varepsilon _{1} = \varepsilon _{2}$.

\section{Energy shifts in the Hydrogen spectrum}

\label{Energy-Levels-Hydrogen}

The optical and Zeeman-type energy shifts for hydrogen can be directly obtained from Eqs. (\ref{DELTAU2}) and (\ref{EnergyShiftsA}), respectively, by taking $i = s = 1/2$ and $Z = 1$. We observe that the Zeeman splitting $E _{\theta}$ vanishes for S states since $g _{\mbox{\scriptsize fs}} ^{(1)} = 0$. Therefore, the lowest lying lines for which the TME becomes manifest are the P states. In the following we discuss the spectroscopic transitions for the $n$P$_{3/2}$ and $n$P$_{1/2}$ lines.

\subsection{Spectroscopy of the $n$P$_{3/2}$ states}

\label{SpectroscopyNP3/2}

\begin{figure}
\begin{center}
\includegraphics[scale=0.38]{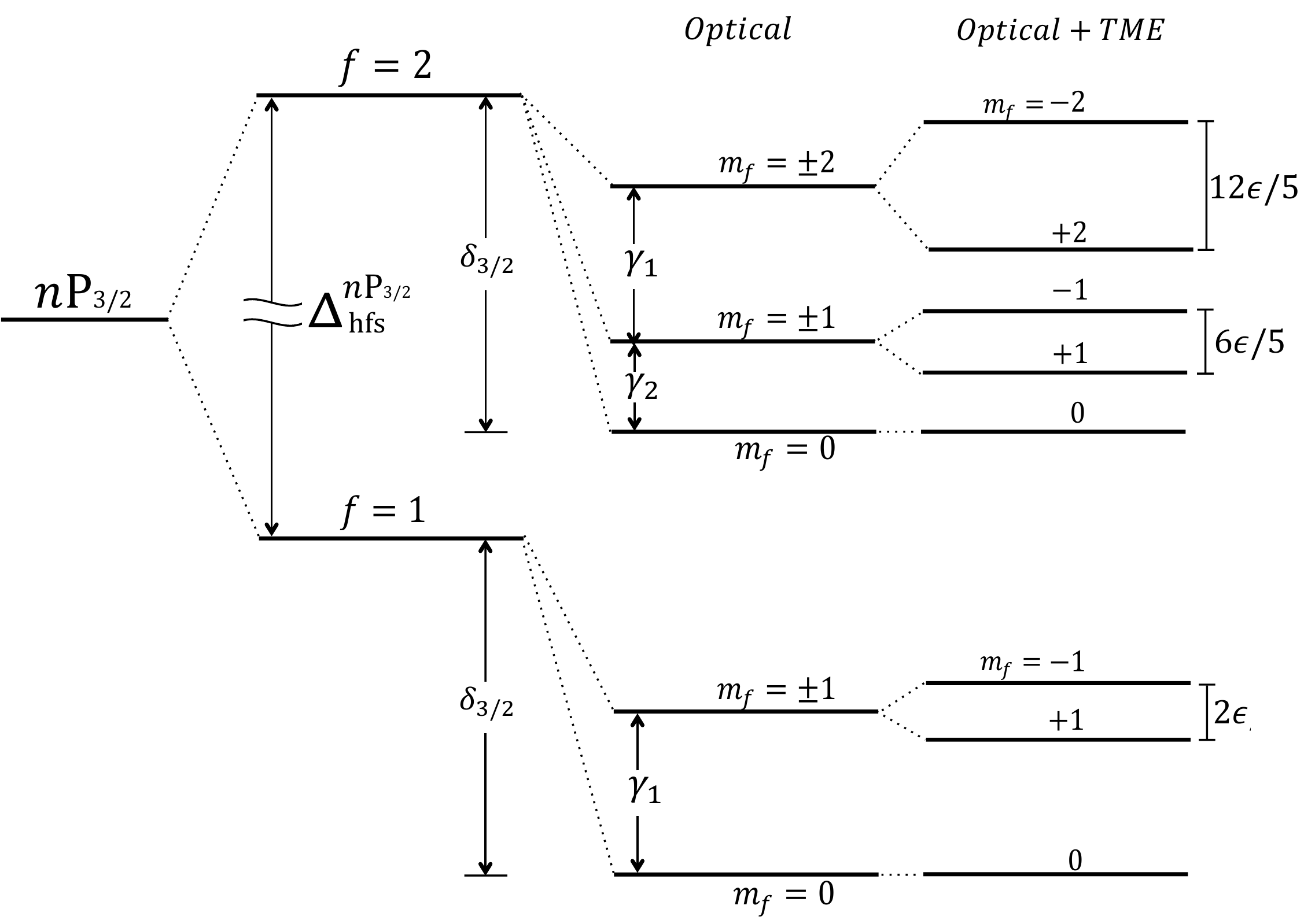}
\end{center}
\caption{Splitting of the $n$P$_{3/2}$ line.}
\label{split2}
\end{figure}

In this particular case we take the set of quantum numbers $\left\lbrace s , \ell , j , i \right\rbrace = \left\lbrace 1/2 , 1 , 3/2 , 1/2 \right\rbrace$, such that the atomic angular momentum can take the values $f = 3/2 \pm ^{\prime} 1/2$. The optical energy shifts then become
\begin{align}
\left< \delta U _{2} \right> ^{n\mbox{\scriptsize P} _{3/2}} _{\pm ^{\prime}} = \frac{\kappa \varepsilon _{1} ^{2} \xi ^{3}}{48} \vert E _{g} \vert n ^{2} (n ^{2} -1) \left( 22 - 2 m _{f} ^{2} \pm ^{\prime} m _{f} ^{2} \right) , \label{OpticalnP3/2}
\end{align}
where the values of $m _{f}$ are restricted by $- f \leq m _{f} \leq f$. The Zeeman-like energy shifts take the form
\begin{align}
\langle\delta V_{\theta}\rangle_{\pm ^{\prime}}^{n\mbox{\scriptsize P} _{3/2}} = \frac{ge \xi ^{2}}{24} E _{g} (4 \mp ^{\prime} 1) m _{f}  . \label{TopologicalnP3/2}
\end{align}
Also, from Eq. (\ref{HyperfineHIon}) one can further obtain that the unperturbed hyperfine energy levels are
\begin{align}
E _{\mbox{\scriptsize hfs}} ^{n\mbox{\scriptsize P} _{3/2}} &= \frac{\Gamma _{3/2}}{n ^{3}} \left[ f (f+1) - 9/2 \right] , \label{HyperfineNP3/2}
\end{align}
where $\Gamma _{3/2} = 1.95 \times 10 ^{-7}$eV. In Fig. \ref{split2} we present the general energy shifts of the $n$P$_{3/2}$ line in two steps. We first observe that the optical contribution (\ref{OpticalnP3/2}) partially breaks the degeneracies of the hyperfine levels, but the  degeneracy of the levels with $m _{f} \neq 0$ is still present. Finally we add the contributions from
$\langle\delta V_{\theta}\rangle_{\pm ^{\prime}}^{n\mbox{\scriptsize P} _{3/2}}$ 
arising from the TME and observe that this effect completely breaks the degeneracy of the hyperfine states $\left| j, f , m _{f} \right> _{\mbox{\scriptsize hfs}}$. The values of the parameters appearing in Fig. \ref{split2} are
\begin{gather}
\delta _{3/2} = 22 (\gamma _{1}/3) = 22 \gamma _{2} = \frac{11}{24} \varepsilon _{1} ^{2} \xi ^{3} \vert \kappa E _{g} \vert n ^{2} (n ^{2} - 1) \notag \\  \epsilon = \frac{5}{24} \vert ge E _{g} \vert \xi ^{2} \quad , \quad \Delta _{\mbox{\scriptsize hfs}} ^{n\mbox{\scriptsize P} _{3/2}} = \frac{4}{n ^{3}} \Gamma _{3/2} . \label{PARAM32} 
\end{gather}

In the following we determine the parameter region where our results are included. To the best of our knowledge the hyperfine splitting of the lines $n$P$_{3/2}$ has not yet been measured and the existing data corresponds to theoretical calculations, which produce the value $\Delta _{\mbox{\scriptsize hfs}} ^{n\mbox{\scriptsize P} _{3/2}} $ together with a theoretical uncertainty $\Delta _{\mbox{\scriptsize tu}} ^{n\mbox{\scriptsize P} _{3/2}}$ \cite{Table1, Table2}. In order for our results to be accessible from a theoretical perspective, we have to identify a range of distances for $b$ satisfying the following conditions: (i) on one hand, $b$ should be such that the additional distance-dependent energy shifts (\ref{OpticalnP3/2}) and (\ref{TopologicalnP3/2}) are larger than the theoretical uncertainty $\Delta _{\mbox{\scriptsize tu}} ^{n\mbox{\scriptsize P} _{3/2}}$, but smaller than the corresponding hyperfine splitting $\Delta _{\mbox{\scriptsize hfs}} ^{n\mbox{\scriptsize P} _{3/2}}$, and (ii) on the other hand, $b$ should be larger than the thickness of the magnetic coating covering the TI surface (to ensure a negligible wave-function overlap) and smaller than the wavelength $\lambda_{\rm C} $ of a typical  hyperfine transition (to ensure the validity of the nonretarded regime). Typically $\lambda _{\rm C} \sim $ m, while a ferromagnetic covering for the TI (such has GdN) can be grown as a thin film of thickness $\sim 6$nm \cite{Grushin-PRL}. Thus, $b \sim \mu $m satisfies the required conditions in practical terms. This choice will restrict the possible values for the remaining parameters in the energy shifts, namely the principal quantum number $n$ and the permittivities $\varepsilon _{1}$ and $\varepsilon _{2}$. Our primary interest here is to accommodate the Zeeman-type splitting $\langle\delta V_{\theta \, \pm ^{\prime}}\rangle^{n\mbox{\scriptsize P}_{3/2}}$, which does not depend on $n$ but depends inversely on the permittivities. Thus low values of $\varepsilon _{1}$ and $\varepsilon _{2}$ favor this contribution. Taking into account the above considerations we find that the recently discovered topological insulator TlBiSe$_{2}$, for which $\theta = \pi$, $\mu = 1$ and $\varepsilon \approx 4$ \cite{TlBiSe2, TlBiSe22, TlBiSe23}, is a good candidate to illustrate our procedure. Assuming that the dielectric medium has also a low permittivity we find that the Zeeman-type energy splitting $2 \epsilon$ is of the order of $10 ^{-12}$eV, while the maximum optical splitting $\gamma _{1} + \gamma _{2}$ becomes of the order of $n ^{2} (n ^{2} - 1) \times 10 ^{-12}$eV for $\varepsilon _{1} \neq \varepsilon _{2}$ and $n ^{2} (n ^{2} - 1) \times 10 ^{-16}$eV for $\varepsilon _{1} = \varepsilon _{2}$. Now, since we have computed the energy shifts from a perturbative perspective, we must guarantee the validity of perturbation theory by imposing the energy splitting $2 \epsilon$ and $\gamma _{1} + \gamma _{2}$ to be at least three orders of magnitude smaller than $\Delta _{\mbox{\scriptsize hfs}} ^{n\mbox{\scriptsize P} _{3/2}}$ thus restricting the possible values for the principal quantum number $n$. Therefore, from (\ref{PARAM32}) we can see that $n=2$ is the best option for the $\varepsilon _{1} \neq \varepsilon _{2}$ case, since $\Delta _{\mbox{\scriptsize hfs}} ^{2 \mbox{\scriptsize P} _{3/2}} = 9.75 \times 10 ^{-8}$eV and $\Delta _{\mbox{\scriptsize tu}} ^{2 \mbox{\scriptsize P} _{3/2}} = 2.47 \times 10 ^{-12}$eV, thus having a range of four orders of magnitud to accomodate the energy shifts. For the case $\varepsilon _{1} = \varepsilon _{2}$ one can choose $n=10$ such that $\gamma _{1} + \gamma _{2} \approx \Delta _{\mbox{\scriptsize hfs}} ^{10 \mbox{\scriptsize P} _{3/2}} \times 10 ^{-3}$, however in this case the energy shifts become smaller than the theoretical uncertainty. Therefore, based on the previous analysis, for definiteness let us consider the atom to be in vacuum in front of the TI TlBiSe$_{2}$ and consider the energy shifts on the $2$P$_{3/2}$ line.
\begin{figure}[tbp]
\begin{center}
\includegraphics[scale=0.33]{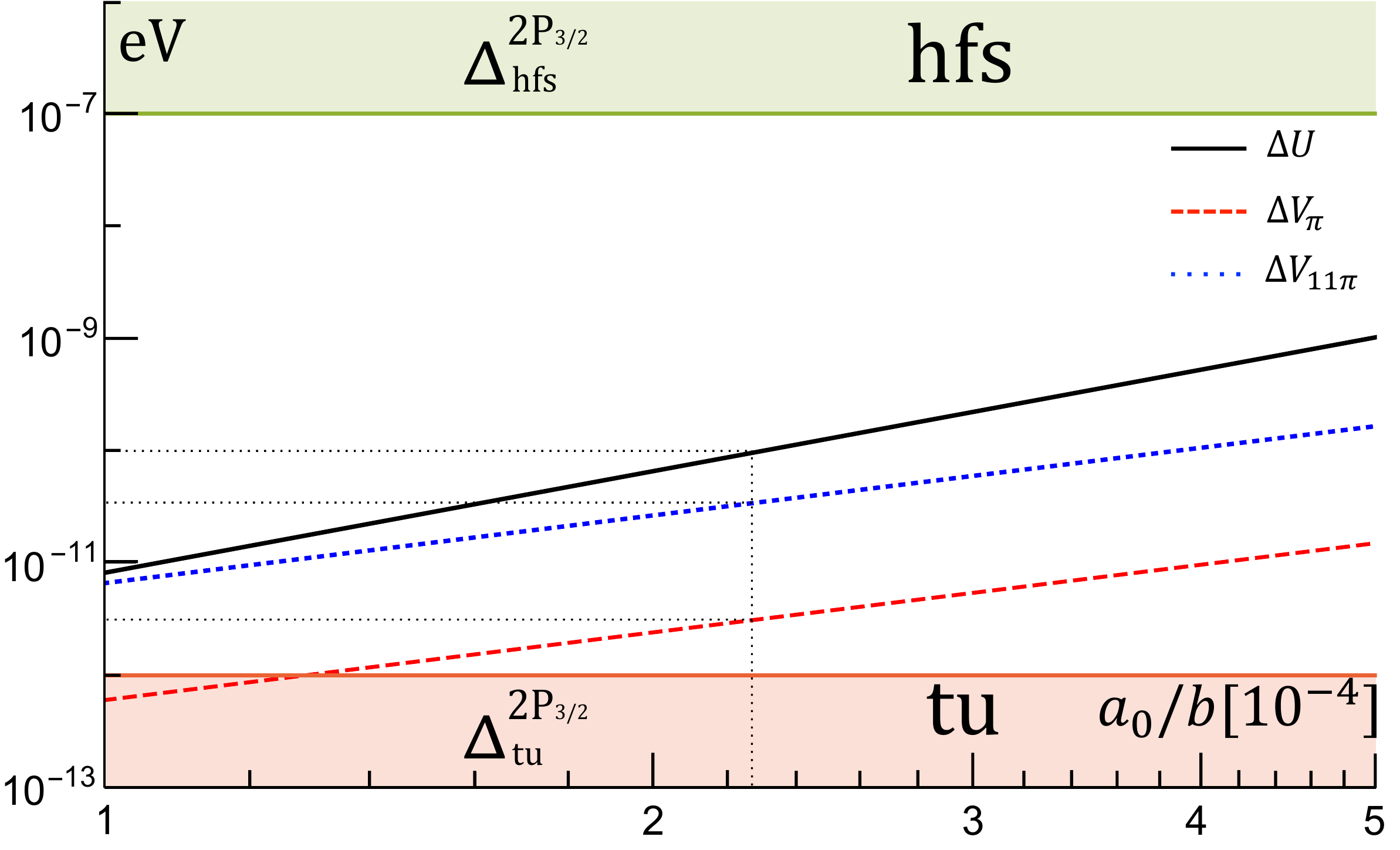}
\end{center}
\caption{{\protect\small Region of accessibility for detecting the TME in hydrogen. The green (upper) and orange (lower) shaded regions are forbidden by the hyperfine splitting and the theoretical uncertainty of the $2$P$_{3/2}$ line, respectively. The selected value of $b$ is $0.23 \mu$m.}}
\label{figure2}
\end{figure}

Using the previously chosen set of parameters we establish the functions $\Delta U \equiv \gamma _{1} + \gamma _{2} = (3/5) \vert E _{g} \vert \xi ^{3}$ and $\Delta V _{\theta} \equiv 2 \epsilon = (1/12) \vert (\theta / \pi) E _{g} \vert \alpha ^{2} \xi ^{2}$ for the maximum optical and Zeeman-type splitting, respectively. By imposing $\Delta U = \Delta _{\mbox{\scriptsize hfs}} ^{2 \mbox{\scriptsize P} _{3/2}} \times 10 ^{-3}$ we determine $b \sim 4337 a _{0} = 0.23 \mu$m, which is a distance perfectly achievable with current experimental techniques. In Fig. \ref{figure2} we present a log-log plot of $\Delta U$ (black line) and $\Delta V _{\theta}$, for two cases: $\theta = \pi$ (dashed red line) and $\theta = 11 \pi$ (dotted blue line). The green (upper) and orange (lower) shaded regions are forbidden by the upper bound $\Delta _{\mbox{\scriptsize hfs}} ^{2 \mbox{\scriptsize P} _{3/2}}$ of the hyperfine structure and by the lower bound arising from the theoretical uncertainty $\Delta _{\mbox{\scriptsize tu}} ^{2 \mbox{\scriptsize P} _{3/2}}$, respectively. We observe that the Zeeman-type contributions, although smaller than the optical one, are bigger than the theoretical uncertainty associated with the determination of the hyperfine splitting of the $2$P$_{3/2}$ state. For $\theta = \pi$ we find $\Delta V _{\pi} = 3.15 \times 10 ^{-12}$eV, while for $\theta = 11 \pi$ we have $\Delta V _{11 \pi} = 3.47 \times 10 ^{-11}$eV. The resulting values of the parameters (\ref{PARAM32}) are $\delta _{3/2} = 133$kHz, $\gamma _{1} = 18$kHz, $\gamma _{2} = 3$kHz and $\epsilon =388$Hz for $\theta = \pi$ and $\epsilon =4.2$kHz for $\theta = 11 \pi$.

\subsection{Spectroscopy of the $n$P${}_{1/2}$ states}
\label{SpectroscopyNP1/2}

\begin{figure}
\begin{center}
\includegraphics[scale=0.38]{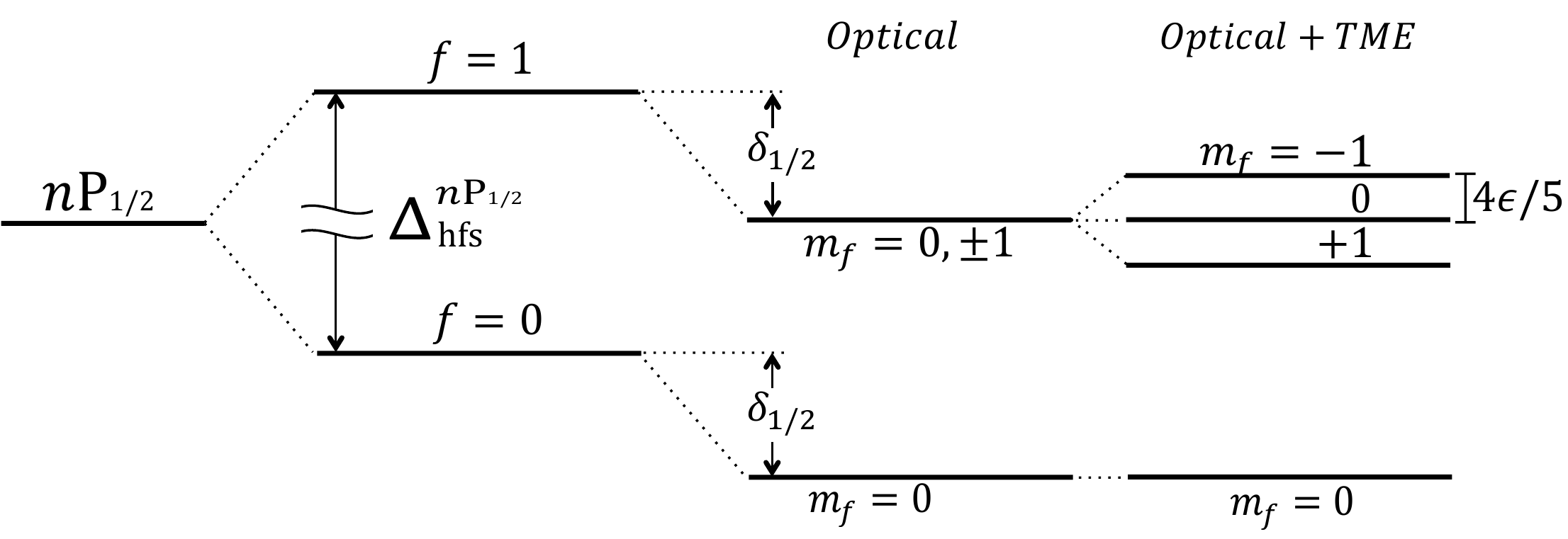}
\end{center}
\caption{Splitting of the $n$P${}_{1/2}$ line.  }
\label{split1}
\end{figure}
Now we consider the set of quantum numbers $\left\lbrace s , \ell , j , i \right\rbrace = \left\lbrace 1/2 , 1 , 1/2 , 1/2 \right\rbrace$, such that the atomic angular momentum can take the values $f = 1/2 \pm ^{\prime} 1/2$. The energy shifts can be obtained directly from Eqs. (\ref{DELTAU2}) and (\ref{EnergyShiftsA}) in a simple fashion. For the optical energy shifts one finds
\begin{align}
\left< \delta U _{2} \right> ^{n\mbox{\scriptsize P} _{1/2}} = \frac{5}{12} \kappa \varepsilon _{1} ^{2} \xi ^{3} \vert E _{g} \vert n ^{2} \left( n ^{2} - 1 \right) , \label{OpticalnP1/2}
\end{align}
which is independent of $f$, while for the Zeeman-type contribution one obtains
\begin{align}
\langle\delta V_{\theta}\rangle_{\pm ^{\prime}}^{n\mbox{\scriptsize P} _{1/2}} =  \frac{1}{6} g e \xi ^{2} E _{g} m _{f} , \label{TopologicalnP1/2}
\end{align}
where the value of $m _{f}$ is restricted by $- f \leq m _{f} \leq f$. Also one can further obtain that the unperturbed hyperfine energy levels are
\begin{align}
E _{\mbox{\scriptsize hfs}} ^{n\mbox{\scriptsize P} _{1/2}} &= \frac{\Gamma _{1/2}}{n ^{3}} \left[ f (f+1) - 3/2 \right] , \label{HyperfineNP1/2}
\end{align}
where $\Gamma _{1/2} = 5 \Gamma _{3/2} = 9.8 \times 10 ^{-7}$eV. In Fig. \ref{split1} we present the energy shifts of the $n$P${}_{1/2}$ line. We observe that the optical energy shift does not break the degeneracy of the $n$P${}_{1/2}$ line, but the Zeeman-type shift does. The values of the parameters appearing in Fig. \ref{split1} are
\begin{align}
\delta _{1/2} = (10/11) \delta _{3/2} \quad , \quad  \Delta _{\mbox{\scriptsize hfs}} ^{n\mbox{\scriptsize P} _{1/2}} = (5/2) \Delta _{\mbox{\scriptsize hfs}} ^{n\mbox{\scriptsize P} _{3/2}} . \label{ParametersNP1/2}
\end{align}
where the parameters in the rhs are the previously defined in Eq. (\ref{PARAM32}). Now we follow the same reasoning of the previous section to analyze the parameter region where our results are included. For distances of the order of $\mu$m we also conclude that low values of the permittivities and low values of the quantum number $n$ favor the Zeeman-type splitting. Thus we consider the atom to be in vacuum in front of the TI TlBiSe$_{2}$ and consider the energy shifts on the $2$P${}_{1/2}$ line. By imposing the maximum optical energy shift $\delta _{1/2}$ to be three orders of magnitud smaller than the hyperfine splitting $\Delta _{\mbox{\scriptsize hfs}} ^{2 \mbox{\scriptsize P} _{1/2}}$ we determine $b \sim 5511 a _{0} = 0.29 \mu$m. The corresponding energies splitting become, $\delta _{1/2} = 2.43 \times 10 ^{-10}$eV and $\epsilon = 192$Hz.

 \section{The Casimir-Polder interaction in the $2$P$_{j}$ line}

\label{Casimir-Polder}

In this section we examine the corrections due to the TME to the standard Casimir-Polder potential in order to determine their possible impact  upon some scattering experiments designed to test the potential \cite{Shimizu, Friedrich}. The general form of the CP potential for a  hydrogen atom as a function of $y = \xi ^{-1}$ is
\begin{align}
V _{\mbox{\scriptsize CP}} \left( y \right) = - \vert E _{g} \vert \left[ \frac{P}{y ^{3}} + (\theta / \pi) m _{f} \frac{Q}{y ^{2}} \right] , \label{CP2P}
\end{align}
with
\begin{align}
P &= \frac{\varepsilon _{1}}{8} \frac{2  (\varepsilon _{2} - \varepsilon _{1}) + \tilde{\theta} ^{2}}{2 (\varepsilon _{2} + \varepsilon _{1}) + \tilde{\theta} ^{2}} \left< \tilde{r} ^{2} \right> _{n \ell} \Sigma _{Z=1} ^{ijf m _{f}} , \notag \\ Q &= \frac{ \alpha ^{2}}{2} \frac{g ^{(Z=1)} _{\mbox{\scriptsize fs}} g _{\mbox{\scriptsize hfs}}}{2 (\varepsilon _{2} + \varepsilon _{1}) + \tilde{\theta} ^{2}} , \label{ParametersCP}
\end{align}
where we have taken $\mu _{1} = \mu _{2} = 1$. Note that the value of the parameters (\ref{ParametersCP}) depend on the specific state under consideration. Also we observe that $Q \geq 0$, while $P \geq 0$ for $\varepsilon _{2} \geq \varepsilon _{1}$ and $P < 0$ for $\varepsilon _{2} < \varepsilon _{1}$. Here we consider $P \geq 0$, which is the physically interesting case.

\begin{figure}[tbp]
\begin{center}
\includegraphics[scale=0.42]{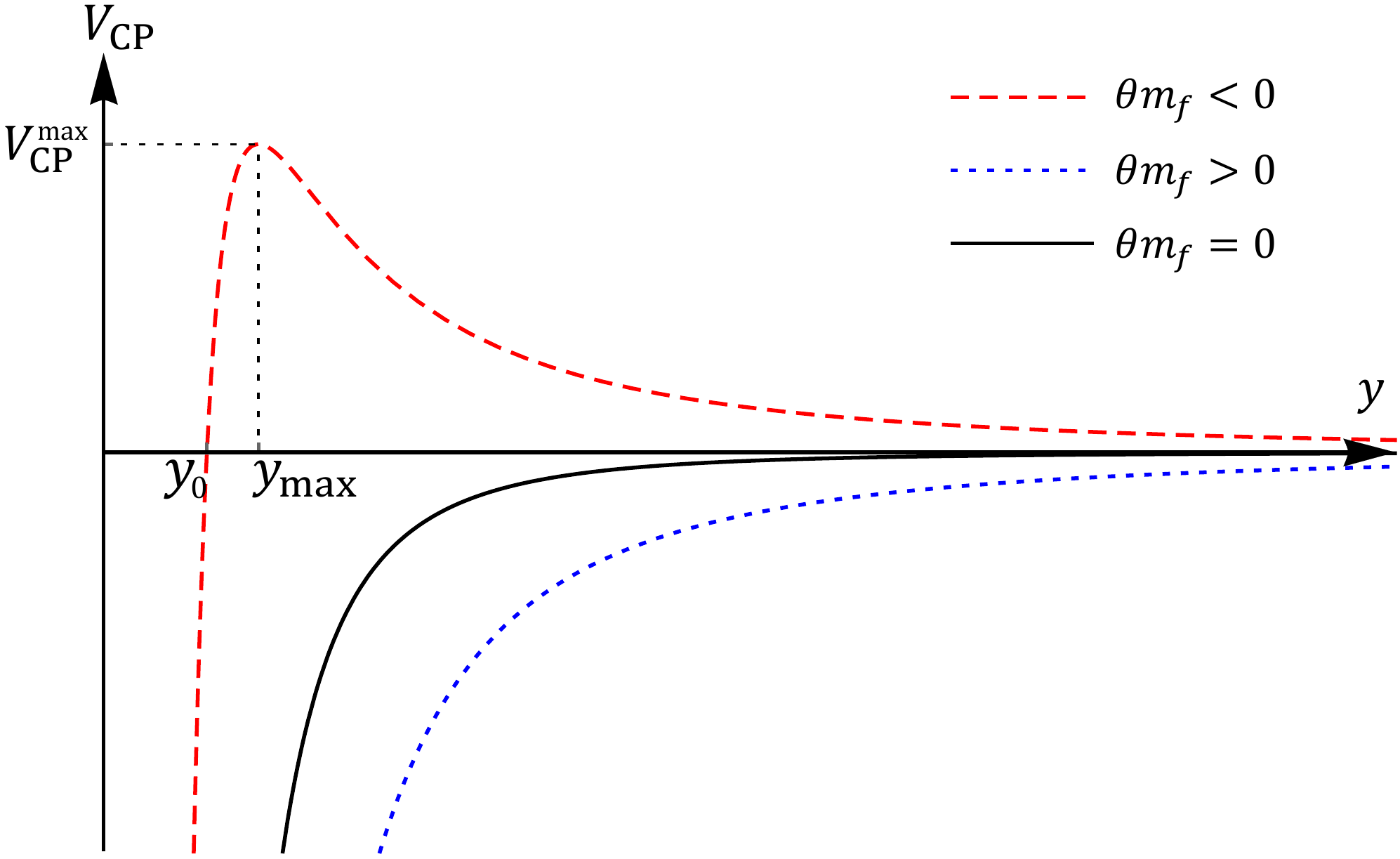}
\end{center}
\caption{{\protect\small Tunability between the attractive and repulsive character of the Casimir-Polder potential $V _{\mbox{\scriptsize CP}} (y)$. }}
\label{CPPOT}
\end{figure}

The magnetoelectric effects arise predominantly from the term proportional to $\theta m _{f} Q$, though there are also corrections in the coefficient $P$. For $\theta = 0$, Eq. (\ref{CP2P}) correctly reduces to the usual attractive CP potential between a hydrogen atom and a dielectric half-space. However, when $\theta \neq 0$ is considered, two interesting cases appear: (i) $ \theta m _{f} > 0$ and (ii) $\theta m _{f} < 0$. In the first case the full CP potential (\ref{CP2P}) retains its original attractive form in the whole range  $0 < b < \infty $, showing only a slight decrease with respect to the usual case. The second case is in principle more interesting because now the second term in Eq. (\ref{CP2P}) is negative; therefore, the potential goes to $- \infty$ when $b$  approaches zero, but tends to zero, from the positive side, when $b\rightarrow \infty$. In fact, one can further show that the CP potential (\ref{CP2P}) has a zero at $b _{0}$, with
\begin{equation}
y _{0} = \frac{b _{0}}{a _{0}} = \vert (\theta / \pi) m _{f} \vert ^{-1} \frac{P}{Q} , \label{YZERO}
\end{equation}
and a positive maximum at $y _{\mbox{\scriptsize max}} = b_{\mbox{\scriptsize max}} / a _{0} = 3 y _{0} / 2$, given by
\begin{equation}
V _{\mbox{\scriptsize CP}} ^{\mbox{\scriptsize max}} = \vert E _{g} \vert \frac{P}{2 y ^{3} _{\mbox{\scriptsize max}}} . \label{MAXCP}
\end{equation}
In other words, for $\theta m _{f} < 0$ the CP potential is attractive in the range $0 < b < b_{\mbox{\scriptsize max}}$ and repulsive when $b_{\mbox{\scriptsize max}} < b$. A generic and  very qualitative form of the CP potential (\ref{CP2P}) is shown in Fig. \ref{CPPOT}. This type of potentials are known in the literature as attractive potential tails and they lead to the phenomena known as quantum reflection, which have attracted great attention in recent years from both the theoretical and experimental sides. In some applications the repulsive contribution to the CP potential is induced by evanescent light above a glass surface; however, in our CP potential (\ref{CP2P}) the repulsive tail is generated by the interaction between the image magnetic monopoles and the atomic angular momentum. Actually, the attractive and repulsive character of the CP potential (\ref{CP2P}) can be tuned by means of the TMEP $\theta$ for a given value of $m _{f}$, where the sign of $\theta$ is determined by the direction of the magnetization of the coating on the surface of the TI.

Now we are interested in making some estimations regarding the position of the maximum $b_{\mbox{\scriptsize max}}$ and the value of the CP potential there $V _{\mbox{\scriptsize CP}} ^{\mbox{\scriptsize max}}$. Let us consider the nuclear charge $Z$, the permittivities $\varepsilon _{1}$ and $\varepsilon _{2}$, and the TMEP $\theta$ as the independent variables which we can control. Because we are working in the nonretarded approximation we have to make sure that the range of applicability $b$ of our estimations is such that $b < \lambda _{\mbox{\scriptsize C}}$,  where $\lambda _{\mbox{\scriptsize C}}$ is a wavelength characteristic of the atomic transitions to be probed and which depends crucially upon the experimental setup. The  values of $\lambda _{\mbox{\scriptsize C}}$  range from $\sim 7.8 \,\mu m$, when dealing with transitions in the $n=1,2,3$ sector, to values of $\sim 7.9 \times 10^7 \, \mu m$, for transitions within the $n=2$ hyperfine sector.

A simple analysis reveals that low principal quantum numbers together with high nuclear charges favor the maximum of the CP potential. Thus, let us restrict to the analysis of the maximally projected $2$P$_{3/2}$ states considered in the previous section, i.e. with $f = 2$ and $m _{f} = -2$ provided $\theta > 0$, or alternatively, with $m _{f} = + 2$ and $\theta < 0$. In this way, from Eqs. (\ref{YZERO}) and (\ref{MAXCP}) we obtain
\begin{align}
y _{\mbox{\scriptsize max}} &= \frac{15 \varepsilon _{1}}{\alpha} \frac{2 (\varepsilon _{2} - \varepsilon _{1}) + \tilde{\theta} ^{2}}{\tilde{\theta}} , \label{YMAXF} \\ V _{\mbox{\scriptsize CP}} ^{\mbox{\scriptsize max}} &= \frac{\vert E _{g} \vert}{1332} \frac{\alpha ^{3}}{\varepsilon _{1} ^{2}} \frac{\tilde{\theta} ^{3}}{[2 (\varepsilon _{2} + \varepsilon _{1}) + \tilde{\theta} ^{2}] [2 (\varepsilon _{2} - \varepsilon _{1}) + \tilde{\theta} ^{2}] ^{2}} . \notag
\end{align}
One can further see that both $y _{\mbox{\scriptsize max}}$ and $ V _{\mbox{\scriptsize CP}} ^{\mbox{\scriptsize max}}$ have critical points at $\tilde{\theta} = \sqrt{2 (\varepsilon _{2} - \varepsilon _{1})}$ and $\tilde{\theta} = \sqrt{-(2 \varepsilon _{1} / 3) + \sqrt{(2 \varepsilon _{1} / 3) ^{2} + 4 (\varepsilon _{2} ^{2} - \varepsilon _{1} ^{2})}}$, respectively. However none of them are physically accessible provided $\varepsilon _{1} \neq \varepsilon _{2}$ since they require very high values of the TEMP. In the following let us discuss the case of a hydrogen atom near the surface of the TI TlBiSe$_{2}$.
\begin{figure}
\subfloat[$b _{\mbox{\scriptsize max}}$ in units of $\mu$m.,\label{YMAXFIG}]{\includegraphics[scale=0.42]{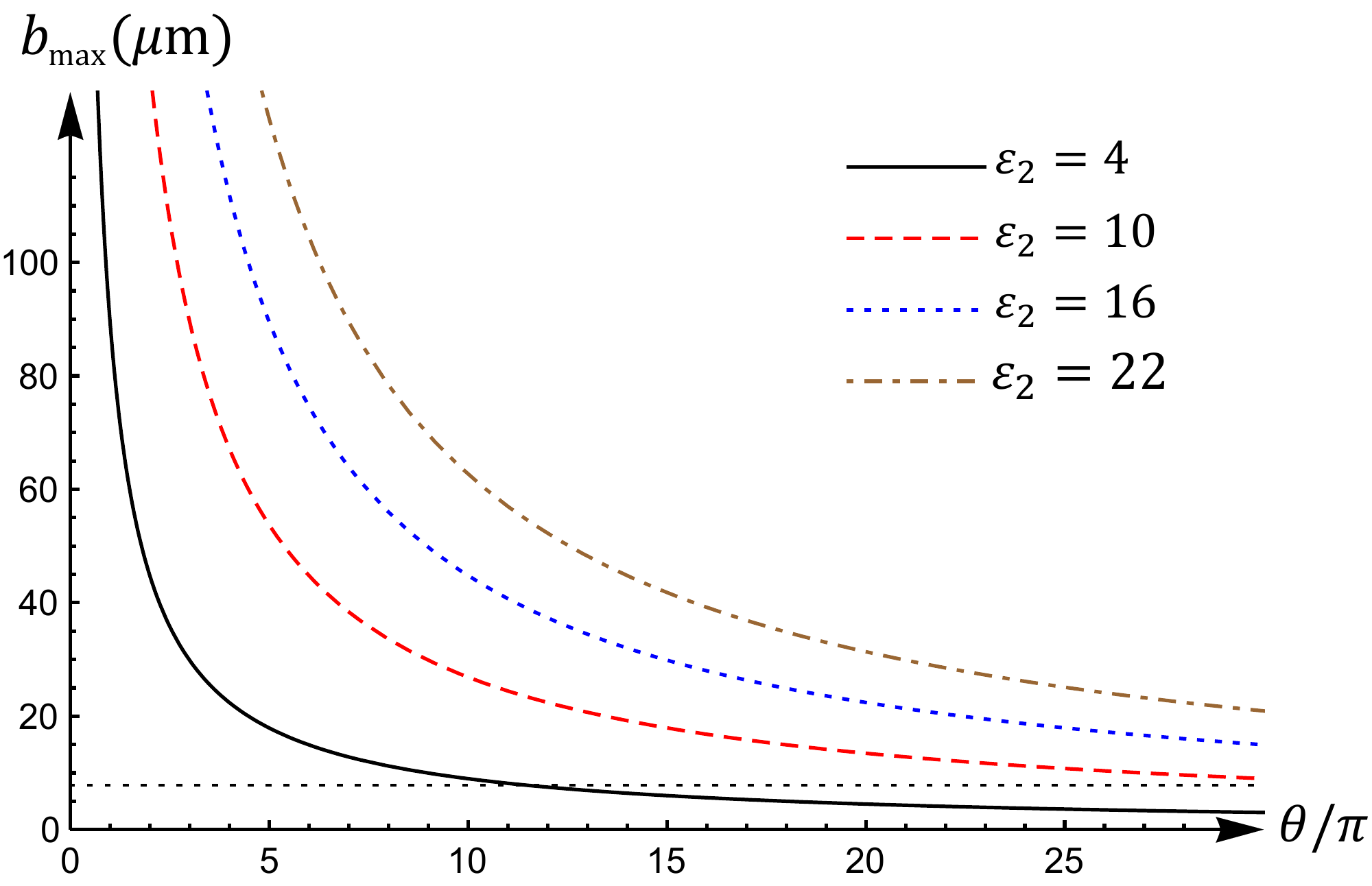}} \\
\subfloat[$V _{\mbox{\scriptsize CP}}^{\mbox{\scriptsize max}}$ in units of Hz.,\label{VMAXFIG}]{\includegraphics[scale=0.4]{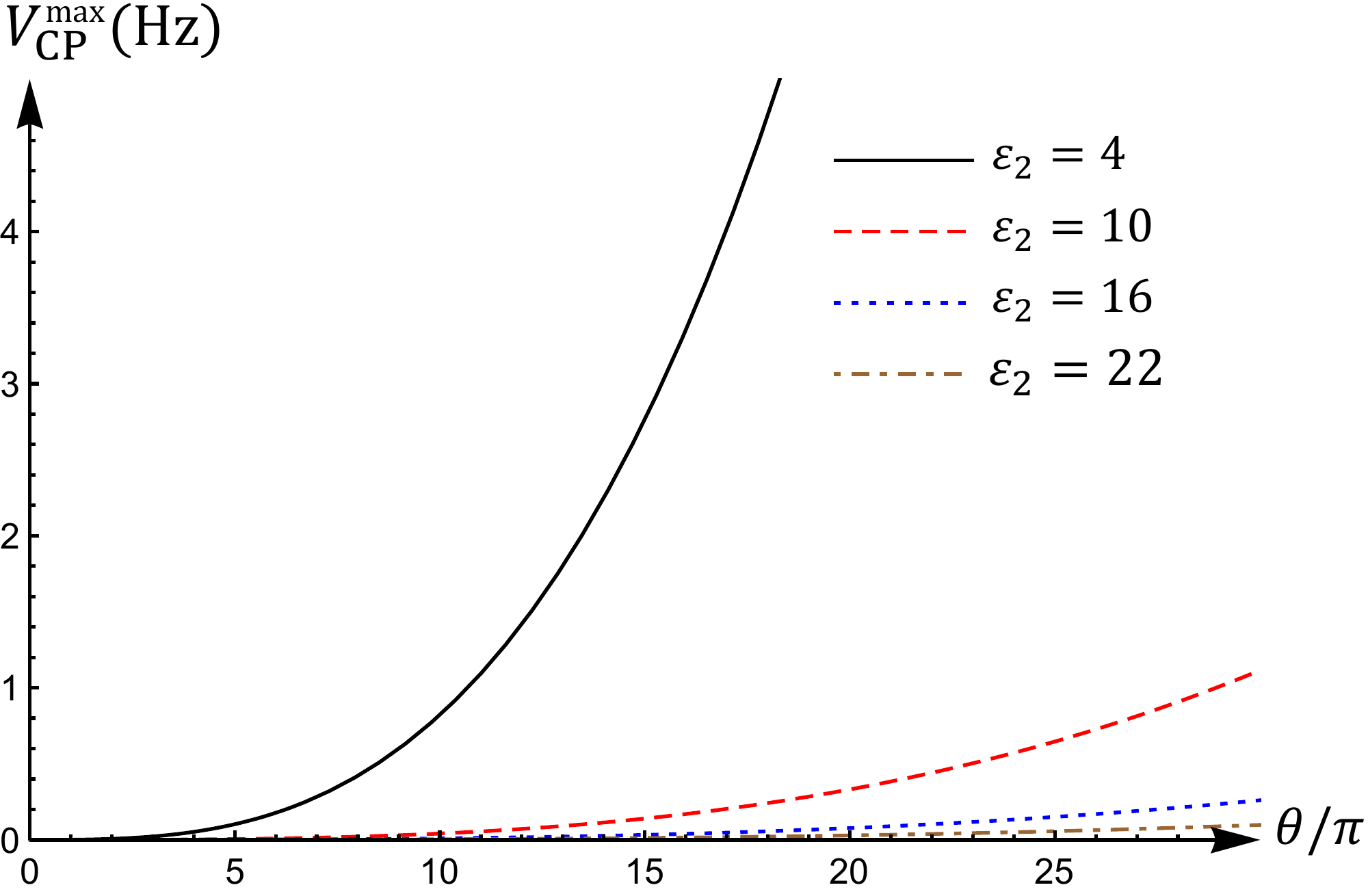}}
\caption{Values of $b_{\mbox{\scriptsize max}}$ and $V _{\mbox{\scriptsize CP}} ^{\mbox{\scriptsize max}}$ for the maximally projected $2$P$_{3/2}$ state as a function of $\theta$ for different values of $\varepsilon _{2}$.}
\label{DensityProfiles}
\end{figure}
In figures {\ref{YMAXFIG}} and \ref{VMAXFIG} we present the $b _{\mbox{\scriptsize max}}$ in units of $\mu$m and the $V _{\mbox{\scriptsize CP}} ^{\mbox{\scriptsize max}}$ in units of Hz, corresponding to the line $2$P$_{3/2}$ with $f = 2$ and $\vert m _{f} \vert = 2$. First we consider the situation where one is probing transitions from the $n = 2$ to the $n = 1$ level, where we must satisfy $b_{\mbox{\scriptsize max}} \ll 7.8 \mu$m, which corresponds to the horizontal dashed line in  Fig. \ref{YMAXFIG}. From the figure we observe that this upper limit  would require larger and larger values of $\theta$ as far as $\varepsilon _{2}$ grows. As emphasized in Refs. \cite{Grushin-PRL} and \cite{Grushin-PRB}, values of $\theta \gtrsim  15 \pi$ induce more general magnetoelectric couplings not included in the effective theory we are considering in Eq. (\ref{Lagrangian}). Thus we take $\theta = 15 \pi$ as an upper value for this parameter. Just requiring $ b_{\mbox{\scriptsize max}}$ to be $ 7.8 \times 10 ^{-1} \mu$m, for example, requires $\varepsilon _{2} = 1.43$ for $\theta = 15 \pi$. Setting the stronger limit of $7.8 \times 10 ^{-2} \mu$m yields $\varepsilon _{2} = 1.038$. The other limiting case is  $\theta = \pi$ which leads to $\varepsilon _{2} = 1.029$ and $\varepsilon _{2} = 1.0029$ respectively. Notice that even the stronger limit  $b _{\mbox{\scriptsize max}} = 7.8 \times 10 ^{-2} \mu$m$=78$nm is larger than the thickness $w$ of the magnetic coating covering the TI surface which is of the order of $6$nm \cite{Grushin-PRL}. Nevertheless, it seems rather unlikely that TI´s with such small values of $\varepsilon _{2}$ are to  be found.  These estimations, together with the energy shifts calculated in the previous sections reinforce the idea that one should probe beyond the hyperfine transitions. In this way, the limiting condition for  the validity of the nonretarded approximation comfortably extends to $ b \ll \lambda_{\rm C} \sim$ m. Because we would like a large value for $V _{\mbox{\scriptsize CP}} ^{\mbox{\scriptsize max}}$, Fig. \ref{VMAXFIG} suggests the choice of small value for $\varepsilon _{2}$ together with a high value for $\theta$. Taking $\varepsilon _{2} = 4$ and $\theta = 15 \pi$ yields $V _{\mbox{\scriptsize CP}} ^{\mbox{\scriptsize max}} \sim 4.2$Hz. Unfortunately this value is about $2 \times 10 ^{-4}$ times smaller than the optical contribution to the splitting of the $f = 2$ level in the $2$P$_{3/2}$ line which is about $21$kHz, as can be seen in  Fig. \ref{split2}. The corresponding location of this maximum is $B_{\rm max}= 5.4 \mu$m. If we substantially increase $V _{\mbox{\scriptsize CP}} ^{\mbox{\scriptsize max}}$ to $1$kHz, just to barely include it in region of accessibility depicted  in Fig. \ref{figure2}, we require the  lower value  $\varepsilon _{2} = 1.3$ with $\theta = 15 \pi$. In this case $B_{\rm max}=0.8\mu$m.

\section{Discussion and conclusions}

\label{Discussion}

In conclusion, we have presented an alternative way to probe the topological magnetoelectric effect (TME) based upon high precision spectroscopy of hydrogenlike ions, including the hydrogen atom, placed at a fixed distance from a planar topological insulator (TI), of which the surface states have been gapped by time-reversal symmetry breaking. We consider the atom to be embedded in a trivial insulator with optical properties $(\varepsilon _{1} , \mu _{1})$; and that the TI is characterized by the set of parameters $(\varepsilon _{2} , \mu _{2} , \theta)$. The coupling between the atomic electron and the image magnetic monopoles produces additional contributions to the Casimir-Polder potential while the ion-TI interaction modiffies the energy shifts in the spectrum, which now became dependent on the ion-surface distance $b$. As expected, we find that the topological contributions are screened by the nontopological ones.

In order to suppress the trivial electrostatic effects we considered the case in which the optical properties of the dielectric medium are comparable with that of the TI, i.e. $\varepsilon _{1} = \varepsilon _{2}$ and $\mu _{1} = \mu _{2} = 1$. In this case, we find a Zeeman-type splitting of the hypefine structure which arises directly from the coupling between the image magnetic monopole fields and the orbital and spin degrees of freedom of the atomic electron. We discussed the lowest lying energy levels where the TME effects become manifest. For hydrogenlike ions ($Z \neq 1$) we find the ground state $1$S$_{1/2}$ to exhibit an energy splitting $\vert \left\langle \delta V _{\theta} \right\rangle \vert \sim 10 $Hz which is a factor $10^{-6}$ smaller than the hyperfine energy level for the $^{3}$He$^{+}$ ion in front of the recently discovered topological insulator TlBiSe$_{2}$. We also find  that circular Rydberg ions can enhance the maximal energy shifts and we determine that the maximum value for the TME correction is $\vert \left\langle \delta V _{\theta} \right\rangle \vert = 1.83 \times 10^{6}$ Hz for the $^{113}$In$^{48+}$ ion. For this improvement to be significant one must probe transitions such that $\Delta n \gg 1$. We demonstrated that the case $\varepsilon _{1} \neq \varepsilon _{2}$ leads to a worse estimations for the maximum energy shifts.

Our analysis of the impact of the TME in circular Rydberg hydrogenlike ions  has been mainly motivated by the recent proposal at NIST of boosting an experimental program for testing theory with one-electron ions in high angular momentum states \cite{Tan}. In fact, more stringent test of theory may be possible  if predictions can be compared with precision frequency measurements in this regime \cite{Jentschura1, Jentschura2}.  As already mentioned, in the case where $\varepsilon _{1} = \varepsilon _{2}$, the optical contribution can be much supressed with respect to that of the TME, which in turn can be of the order of the hyperfine structure energy shifts. Previous measurement of the hyperfine splitting in the ground state of hydrogenlike $^{209}$Bi$^{82+}$ in the optical regime was  reported some time ago in \cite{Klaft}. Thus, having in mind the NIST proposal one might hope that  new techniques in spectroscopy might be able to incorporate higher angular momentum states and also  to integrate  the optical, terahertz and radio-frequency domains \cite{Marian}.

Section \ref{Energy-Levels-Hydrogen} was devoted to the analysis of the interaction between a hydrogen atom ($Z = 1$) in vacuum and the TI TlBiSe$_{2}$. In this case we find that the Zeeman-type splitting is present only in the lines with nonzero angular momentum ($\ell \neq 0$) and is favored by high values of the TMEP $\theta$. We considered such effects on the $2$P$_{3/2}$ and $2$P$_{1/2}$ lines of hydrogen obtaining the value $b = 0.23 \,\mu$m for the effect to be within the theoretical uncertainties in the corresponding parameters of the line. The parameter $\epsilon$, which measures the Zeeman-type energy splitting of the hyperfine structure, as shown in Figs. \ref{split2} and \ref{split1}, is found to be $\epsilon = 388$Hz for $\theta = \pi$, but its value becomes $\epsilon = 4.2$kHz for $\theta = 11 \pi$. We have also discovered an interesting characteristic in the Casimir-Polder potential of the $2$P$_{j}$ lines, which is the tunability between the attractive and repulsive character of the CP interaction. We find that, for $\theta m _{f} > 0$, the CP potential retains its usual attractive form, while for $\theta m _{f} < 0$ it acquires a positive maximum $V _{\rm CP max}$ located at a distance $B_{\max}$, thus implying the CP potential turns out to be  repulsive  for distances $ b > b_{\mbox{\scriptsize max}}$. This is consistent with previous calculations which show that Casimir forces can be repulsive if they involve magnetic moments couplings \cite{Skagerstam,Boyer}. In a similar fashion, it was recently shown that the dynamical properties of the atomic electron can be tuned with the TMEP $\theta$ \cite{Martin-Chan}. For the TI TlBiSe$_{2}$ we obtain $V_{\rm CP max} = 4.2$ Hz, which is $10^{-2}$ smaller than the theoretical uncertainty in the splitting of the $2$P$_{3/2}$ line. This maximum is located at $b _{\mbox{\scriptsize max}} = 5.4 \mu$m. If we substantially increase $V _{\rm CP max}$ to $1$kHz, just to barely include it in region of accessibility depicted  in Fig. \ref{figure2}, we require the rather low value  $\varepsilon = 1.3$ with $\theta = 15 \pi$. In this  case the maximum is located at $b _{\mbox{\scriptsize max}} = 0.8 \mu$m. As shown in Fig. \ref{VMAXFIG} higher values of $V _{\rm CP max}$ can be obtained from low $\varepsilon$ TIs together with high values of $\theta$. Nevertheless, the latter condition demands the inclusion of additional magnetoelectric effects not considered  in our model.

In our work we have assumed that the magnetic coating has no effect on the energy shifts. However the ferromagnet makes a magnetic field which in turn will induce a Zeeman splitting, and thus it is necessary to distinguish between these two contributions in order to measure the topological contribution (\ref{EnergyShiftsA}). In the present case, the magnetic field is sourced by the magnetization $\textbf{M} = M \,\hat{\textbf{e}} _{z}$ of the coating, along the symmetry axis, and we can estimate it as that produced by a magnetic dipole $\textbf{m} = M V \hat{\textbf{e}} _{z}$, where $V$ is the volume of the coating. For a fixed ion-surface distance, this yields to a total Zeeman energy splitting of the form
\begin{align}
\epsilon (M) = a M + \mbox{sgn} (M) \, \epsilon _{\mbox{\scriptsize topo}} ,
\end{align}
where $a$ is a constant. The first term corresponds to the energy shifts due to the magnetic coating, while the second term corresponds to the topological contribution $\left\langle \delta V _{\theta} \right\rangle$ given by Eq. (\ref{EnergyShiftsA}) together with the fact that the sign of $\theta$ is defined by the direction of the magnetization. Consequently, the topological contribution can be obtained by measuring $\epsilon (M)$ at different magnetizations $M$ and extracting the linear extrapolation of $\epsilon (M)$ as $M \rightarrow 0 ^{+}$.

\acknowledgments We acknowledge helpful discussions with R. J\'{a}uregui, J. Jim\'{e}nez-Mier, D. Sahag\'{u}n and A. Cortijo. We thank the referee for his/her comments and suggestions which have substantially improved  the scope of this work. This work is supported in part by Project No. IN104815 from Direcci\'{o}n General Asuntos del Personal Acad\'{e}mico (Universidad Nacional Aut\'{o}noma de M\'{e}xico) and CONACyT (M\'{e}xico), Project No. 237503.

\end{document}